\documentclass[conference]{IEEEtran}
\IEEEoverridecommandlockouts
\usepackage{cite}
\usepackage{amsmath,amssymb,amsfonts}
\usepackage{graphicx}
\usepackage{textcomp}
\usepackage{xcolor}
\usepackage{makecell}
\usepackage{changepage}
\usepackage{tabularx}
\usepackage{array}
\usepackage{placeins}
\usepackage{float}
\usepackage{booktabs}
\usepackage{multirow}
\usepackage{graphicx}
\usepackage{subfigure}   
\usepackage[ruled,vlined]{algorithm2e}
\usepackage{pgfplots}
\usepackage{subfigure}
\usepackage{xcolor}
\pgfplotsset{compat=1.18}
\usepackage[T1]{fontenc}

\def\BibTeX{{\rm B\kern-.05em{\sc i\kern-.025em b}\kern-.08em
    T\kern-.1667em\lower.7ex\hbox{E}\kern-.125emX}}

\SetKwFunction{Com}{\textsf{Com}}
\SetKwFunction{ZKAgg}{\textsf{ZK$_{\text{Agg}}$}}
\SetKwFunction{ZKCli}{\textsf{ZK$_{\text{Cli}}$}}
\SetKwFunction{ZKInf}{\textsf{ZK$_\infty$}}
\SetKwFunction{ZKTwo}{\textsf{ZK$_2$}}
\SetKwFunction{ZKCos}{\textsf{ZK$_{\text{cos}}$}}
\SetKwFunction{Verify}{\textsf{Verify}}

\begin{document}

\title{ 
Keyed Provenance Watermarking with Complementary Lattice-Based Secure Aggregation for Federated Learning
}





\author{
\IEEEauthorblockN{Xinyun Liu${^a}$\IEEEauthorrefmark{1}, Zhi Lu${^b}$\IEEEauthorrefmark{1}, Yu Chen${^c}$, Ronghua Xu${^a}$}

\IEEEauthorblockA{
$^{a}$Department of Applied Computing, Michigan Technological University, Houghton, MI 49931, USA\\ $^{b}$School of Computing, National University of Singapore, Singapore, 119077, Singapore\\ $^{c}$Department of Electrical and Computer Engineering, Binghamton University, Binghamton, NY 13902, USA\\
\{xinyunl, ronghuax\}@mtu.edu, luzhi@ieee.org, ychen@binghamton.edu
}

\thanks{\IEEEauthorrefmark{1} Xinyun Liu and Zhi Lu contributed equally to this work.}

}

\maketitle

\begin{abstract}

Federated learning (FL) is vulnerable to multi-level attacks. However, existing methods address them separately, leaving FL exposed to data leakage, unauthorized reuse, and malicious gradient manipulation.
In this work, we propose an FL framework that couples keyed context-provenance watermarking with verifiable lattice-based secure aggregation of Real-World Anchored Watermarking and Lattice-Based Zero-Knowledge Secure Aggregation.
At the data layer, we propose a Kerckhoffs-compliant scheme that utilizes Physical Anchor Metadata (PAM) to ensure data provenance.
PAM is defined as a context-provenance token derived from trusted infrastructure data (time, location, and server ID) and then subjected to a keyed HMAC-SHA-256 transformation to produce a watermark payload that cannot be generated without the client's secret key. 
We further design FMGAN, a GAN-based robust image watermarking framework that embeds this transformed payload using a feature fusion module and a Mamba-guided linear attention mechanism.
At the computation layer, we adopt a lattice-based zero-knowledge secure aggregation (LZKSA) protocol that verifies key correctness, $L_{\infty}/L_{2}$ norm bounds, and cosine similarity constraints over committed gradients without revealing private updates. The RLWE-based design guarantees post-quantum security.
Extensive experiments validate the complementary protection of the two layers under composite attack scenarios.
To our knowledge, no prior verification workflow has jointly evaluated both layers in a hybrid, end-to-end trustworthy FL framework.





\end{abstract}


\section{Introduction}
\label{sec:Introduction}

In modern AI ecosystems, Federated learning (FL) constitutes a foundational paradigm for privacy-preserving model training across distributed data silos, seeing rapid integration into safety-critical sectors such as medical imaging \cite{ma2024model}, finance \cite{chatterjee2023federated}, autonomous driving \cite{li2021privacy}, and industrial inspection \cite{liu2025bewsat}.
As FL moves from controlled laboratory settings to complex real-world deployments, establishing the authenticity and verifiable provenance of digital assets has emerged as a paramount security frontier.
Despite an expansion of research in FL security, a fundamental systemic gap remains unresolved: \textbf{existing FL defense mechanisms address either data-layer provenance or gradient-aggregation security in isolation, but never both simultaneously}.
This gap leaves FL systems vulnerable to attack surfaces, including data leakage, unauthorized dataset usage, and malicious model manipulation \cite{liu2025securing,liu2025vulnerability}.



Digital watermarking offers a promising solution to this data-layer challenge by embedding a persistent watermark directly into the media payload, enabling robust tamper detection and asset attribution \cite{liu2024decentralized}.
Nonetheless, existing watermarking methodologies exhibit critical limitations.
Deep learning-based approaches \cite{jia2021mbrs,ahmadi2020redmark} resist specific distortions like noise or JPEG compression but generalize poorly to real-world perturbations and suffer from a brittle robustness-imperceptibility trade-off.
Conversely, hybrid frequency-domain methods \cite{hamidi2021hybrid} withstand geometric and signal-processing attacks but require tedious manual parameter tuning, hindering deployment in distributed environments.
Finally, metadata-based schemes \cite{rosenthol2022c2pa} fail as a secure root of trust; these automated tokens are easily modified or entirely stripped during transmission across platforms.

Secure Aggregation (SA) has been adopted to address the model-layer challenge, in which individual gradients are encrypted, preventing the aggregator from inspecting or filtering malicious updates \cite{zhu2021privacy,eltaras2023efficient}.
Classical secure aggregation protocols \cite{bell2020secure} protect gradient privacy by ensuring that the server observes only aggregated updates, at the expense of verifiability and the ability to inspect anomalies.
While recent Zero-Knowledge Proof (ZKP) frameworks \cite{bell2023acorn,lycklama2023rofl} restore verifiability, general-purpose SNARK or Bulletproof constructions incur prohibitive computational overhead.
Furthermore, their reliance on pre-quantum cryptographic assumptions conflicts with emerging post-quantum standards like FIPS 203  \cite{NIST-FIPS203} and FIPS 204 \cite{NIST-FIPS204}.

To address the limitations identified in both data security and gradient aggregation security, we propose a dual-layer security framework that provides end-to-end protection for federated learning.
At the data layer, we introduce a Kerckhoffs-compliant watermarking scheme where a physical anchor metadata (PAM) is derived from trusted ENF infrastructure and then undergoes a keyed HMAC-SHA256 transformation to produce a watermark payload that is infeasible to generate without the secret key. 
This securely generated watermark is then embedded in images by FMGAN, a GAN-based framework that utilizes feature fusion and Mamba-guided linear attention.
It is \textbf{naturally suited for distributed learning scenarios} since each node possesses its own distinctive PAM, which acts as a context-provenance token synthesized from a multi-dimensional set of environmental identifiers: time, location, and ENF server ID, retrieved directly from a trusted physical anchor infrastructure, and subsequently cryptographically transformed to ensure generation-unforgeability under key secrecy. 
At the computation layer, we adopt a LZKSA protocol to provide cryptographically verifiable correctness of model updates. Through zero-knowledge proofs, each client demonstrates that its committed gradients satisfy prescribed norm constraints, cosine similarity bounds, and key-correctness requirements, without revealing the gradients themselves. The RLWE-based construction ensures post-quantum security while incurring substantially lower overhead than general-purpose ZKP systems such as SNARKs \cite{ben2014succinct} or Bulletproofs \cite{bunz2018bulletproofs}.
To achieve trustworthy FL, we make the following key contributions:
\begin{itemize}
    \item To the best of our knowledge, this is the first work to construct a hybrid, end-to-end trustworthy FL framework that systematically integrates the data-security watermarking layer and the lattice-based zero-knowledge secure aggregation layer in a joint verification workflow.

    \item  We propose a keyed PAM watermarking scheme whose payloads are existentially unforgeable to generate without the secret key, even under full exposure of the algorithm and public parameters. 

    

    \item We design FMGAN, a GAN-based watermarking framework with a feature fusion module and a Mamba-guided linear attention mechanism for embedding cryptographically protected PAM-derived watermarks into images.
    
        
    \item Building on the LZKSA protocol, we present a streamlined, deployment-oriented integration that adapts lattice-based zero-knowledge aggregation to practical federated learning workflows, while preserving verification guarantees and reducing system complexity.

    \item We implement and evaluate a complete end-to-end federated learning pipeline, demonstrating improved robustness against \textbf{dataset leakage, unauthorized dataset usage, and gradient manipulation}, while highlighting the complementary strengths of PAM watermarking and lattice-based secure aggregation.
\end{itemize}

\section{Background and Related Work}
\label{sec:Fundamentals}



\subsection{Image Authentication and Watermarking}
Image authentication approaches are broadly categorized into passive and active techniques.
Passive methods detect tampering solely by analyzing statistical irregularities or forensic artifacts inherent to the digital content \cite{wan2022comprehensive}.
In converse, active techniques embed verifiable data directly into the spatial or frequency domains during digitization \cite{liu2025securing}.
This category includes cryptography, digital signatures, and watermarking for intellectual property protection and content owner identification \cite{lederer2023identifying}.
Among active strategies, digital watermarking provides a distinct advantage by being intrinsically bound to the media payload, ensuring tamper detection and asset attribution even if external metadata is stripped \cite{yang2024gaussian,liu2024decentralized}.
While watermarking spans fragile, semi-fragile, and robust paradigms, robust watermarking is uniquely suited for data ownership verification, as it guarantees payload detectability against both common signal-processing distortions and deliberate adversarial attacks \cite{chen2025imprints}.

\subsection{GAN-based Watermarking methods}

Compared with traditional CNN-based image watermarking techniques, GAN-based watermarking methods have demonstrated superior performance in terms of robustness, invisibility, and adaptability. By leveraging adversarial training, these methods learn to embed watermarks in a data-driven manner, enabling the generator to produce watermarked images that are visually indistinguishable from the originals while remaining resilient to a wide range of distortions. This paradigm has significantly advanced the practicality of robust watermarking in real-world multimedia applications.
Early representative work such as ARWGAN \cite{huang2023arwgan} integrates adversarial learning with encoder–decoder architectures to enhance both concealment quality and robustness. Nevertheless, its reliance on relatively complex network structures introduces higher computational overhead.
MBRS \cite{jia2021mbrs} introduces a mini-batch-based random selection strategy, where the noise layer is randomly chosen from real JPEG compression, simulated JPEG compression, or a noise-free layer for each mini-batch. However, its robustness against other attacks such as Gaussian and median filtering shows limited improvement, and the approach incurs significantly increased training time while struggling to minimize embedding distortion.
SepMark \cite{wu2023sepmark} proposes a multi-embedding framework that jointly supports deepfake detection and source attribution. A single encoder embeds both fragile and robust watermarks, while two separable decoders enable independent extraction for tamper detection and source tracking. Nevertheless, it relies on specific design assumptions and experimental configurations, which may limit its generalization to more diverse scenarios.
Similarly, EditGuard \cite{zhang2024editguard} introduces a unified framework for copyright protection and tamper-agnostic localization by embedding dual watermarks, one for localization and one for ownership verification, allowing accurate recovery even under complex editing operations. However, its design relies on carefully structured editing-aware training and optimization, which introduces additional system complexity.

\subsection{Digital Multimedia Forensics Using ENF Analysis}

The Electric Network Frequency (ENF) represents the nominal mains frequency (50/60 Hz) whose real-time fluctuations mirror grid-wide power load variations.
Because these unique spatio-temporal traces are inadvertently imprinted onto audio or video signals during recording, they serve as a distinct spatio-temporal fingerprint for multimedia authentication and tamper detection.
Grigoras \cite{grigoras2005digital} first demonstrated that ENF fluctuations in audio could provide definitive temporal verification.
This concept was later extended by Cooper et al. \cite{cooper2010forensic} to video by isolating mains-powered illumination flicker.
In social media contexts, existing frameworks like DEMA \cite{nagothu2023dema} and real-world anchors \cite{xu2025detecting, xu2025anchormark} reconstruct ground-truth grid profiles to audit digital asset integrity. Unlike traditional software-generated metadata (e.g., device settings, locations), which is easily modified or stripped during transmission, ENF signals are bound to the physical environment—making them virtually impossible to forge and ideal for reliable leakage detection and data provenance.

\subsection{Secure Aggregation and Zero-Knowledge Proofs in FL}



Although federated learning (FL) restricts raw data on-device, local updates remain vulnerable to inference attacks. Secure Aggregation (SA) mitigates this risk by ensuring the server only observes the aggregated sum of client updates, utilizing pairwise masking and dropout-resilient secret sharing to handle client unreliability at scale \cite{bonawitz2017practical,bell2020secure,stevens2022lightsecagg}.
However, SA exclusively protects \emph{confidentiality} and cannot prevent clients from submitting malformed or adversarial updates.
Zero-Knowledge Proofs (ZKPs) complement SA by enabling clients to non-interactively prove update validity and protocol compliance (e.g., bounded norms, correct execution of a training step, or well-formed encrypted submissions) without exposing private parameters. In practice, ZKP-based FL systems prioritize optimized circuit designs and lightweight statements to reconcile the inherent trade-off between rigorous integrity guarantees and client-side computational overhead.

\section{Problem Statement and Threat Model}

\subsection{System Overview}
We consider a standard cross-device FL setting consisting of a central server (aggregator) and a set of distributed clients. Each client holds a private local dataset and performs on-device training to compute model updates (e.g., gradients or weight differences), which are then sent to the server for aggregation. To preserve data privacy, the server does not have access to raw training data and relies on secure aggregation to compute the global model update.
In this work, we extend the conventional FL pipeline by introducing a dual-layer security architecture. At the data layer, each client embeds a cryptographically PAM watermark as an ENF-grounded provenance token into its local images before training.
At the computation layer, clients participate in a LZKSA protocol, providing cryptographic proofs that their submitted gradients are well-formed and comply with predefined constraints. The server verifies these proofs before aggregating updates, ensuring that only legitimate contributions influence the global model.

\subsection{Security Goals}

Our objective is to establish end-to-end trust in FL under strong and realistic adversarial assumptions. 
The security objectives consist of two complementary components: data authenticity and computation integrity.
At the data layer, the goal is to ensure that each client’s training data is authentic and resistant to forgery or unauthorized reuse. An adversary who knows the watermarking architecture, embedding procedure, and PAM, but does not possess the client’s secret key, should not be able to generate a valid watermark with non-negligible probability. 
The embedded watermark must additionally remain robust under common signal-processing operations and adversarial perturbations.
At the computation layer, the objective is to ensure that model updates are well-formed and compliant with the aggregation protocol without revealing private gradients. Clients must prove correctness of masking keys and adherence to predefined norm and directional constraints. These properties are verified using lattice-based zero-knowledge proofs over RLWE commitments, providing post-quantum security while preserving gradient privacy.



\subsection{Threat Model}

At the data layer, adversaries aim to compromise client-side dataset integrity, authenticity, or provenance via device compromise or insider threats, distributing manipulated data to disrupt collaborative training or sabotage downstream forensic analysis. Regarding adversary capabilities: (1) the attacker possesses full knowledge of the watermarking architecture, embedding/extraction procedures, and public parameters, remaining blind only to the private key driving the HMAC-PAM transformation (with key generation and distribution assumed secure); and (2) a external trusted authority manages and validates the ENF data, preventing the adversary from arbitrarily modifying ground-truth provenance records.
At the gradient and computation layer, we consider malicious FL clients that attempt to compromise the training process by submitting manipulated model updates. These adversaries may execute poisoning or backdoor attacks by deliberately scaling, reversing, randomizing, or adversarially crafting local gradients to disrupt convergence or degrade performance. Additionally, they may deviate from the secure aggregation protocol by uploading malformed payloads or encrypting updates with incorrect cryptographic keys. To mitigate these threats, the server enforces strict input validation, accepting only protocol-compliant updates accompanied by valid cryptographic proofs while rejecting all altered or malformed inputs.



\section{Proposed Method}

\begin{figure}[htbp]
   \begin{center}
    \includegraphics[trim=280 180 250 220, clip, width=0.49\textwidth]{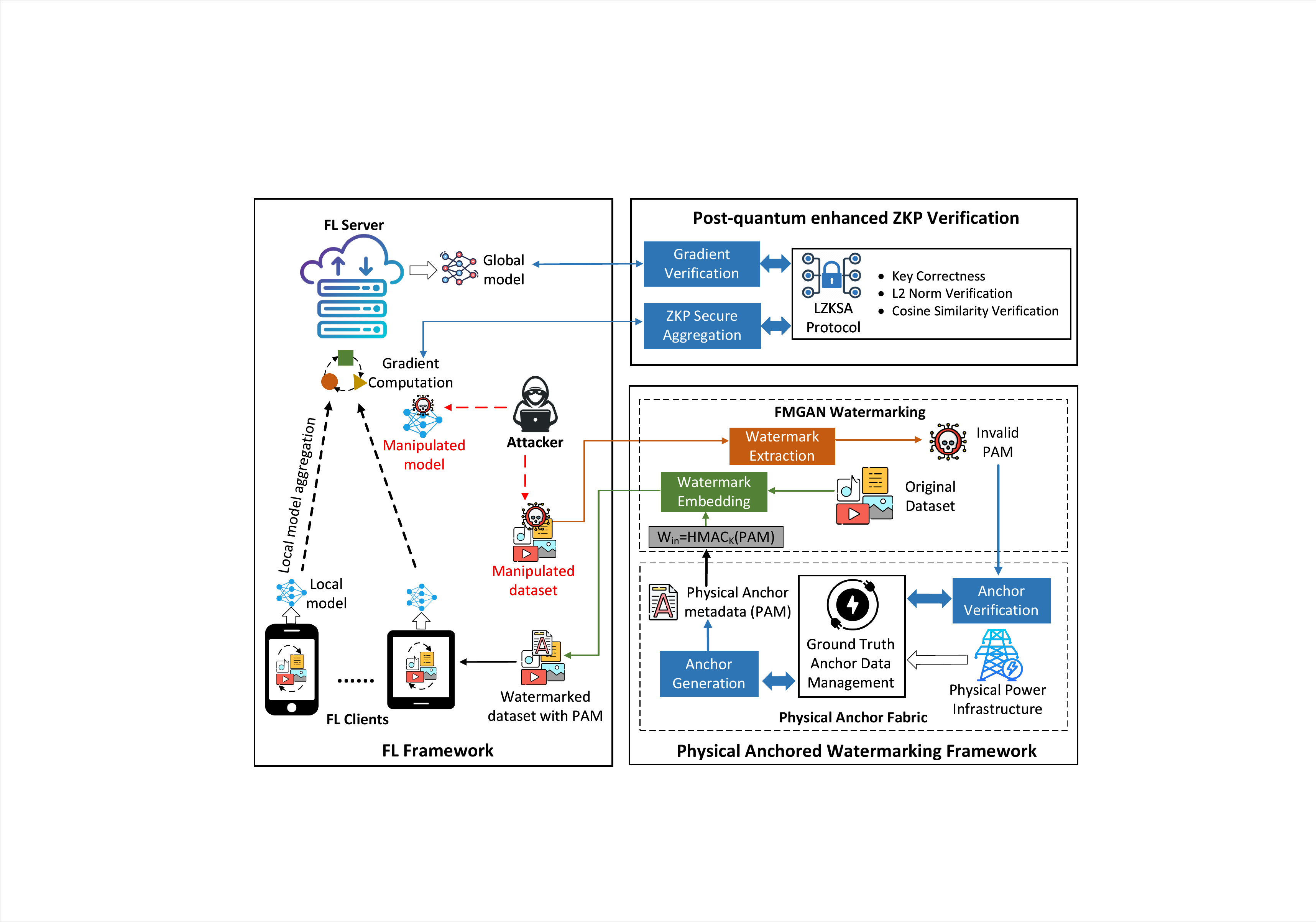}
    \caption { \label{fig:system_design} System Overview.  } 
   \end{center}
    \vspace{-20 pt}
\end{figure}

We propose a dual-layer security framework for FL that jointly enforces data authenticity and computation integrity throughout the training pipeline. As illustrated in Figure \ref{fig:system_design}, the framework integrates a Data-Security Layer based on real-world keyed PAM watermarking with a Computation-Security Layer built upon lattice-based zero-knowledge secure aggregation (LZKSA). By operating on different yet complementary security targets, local training data and aggregated model updates, the two layers collectively establish an end-to-end trust chain from data acquisition to global model updates.

\subsection{Keyed PAM Watermarking}


Our real-world keyed PAM watermarking framework consists of two sub-frameworks: i) a physical anchor fabric that manages ground-truth references and provides PAM generation and verification; ii) an FMGAN-based watermarking framework that embeds cryptographically protected PAM-derived payloads into images and extracts them during the verification stage, as illustrated in Figure~\ref{fig:system_design}.

\begin{figure}[htbp]
   \begin{center}
    \includegraphics[trim=300 180 300 180, clip, width=0.49\textwidth]{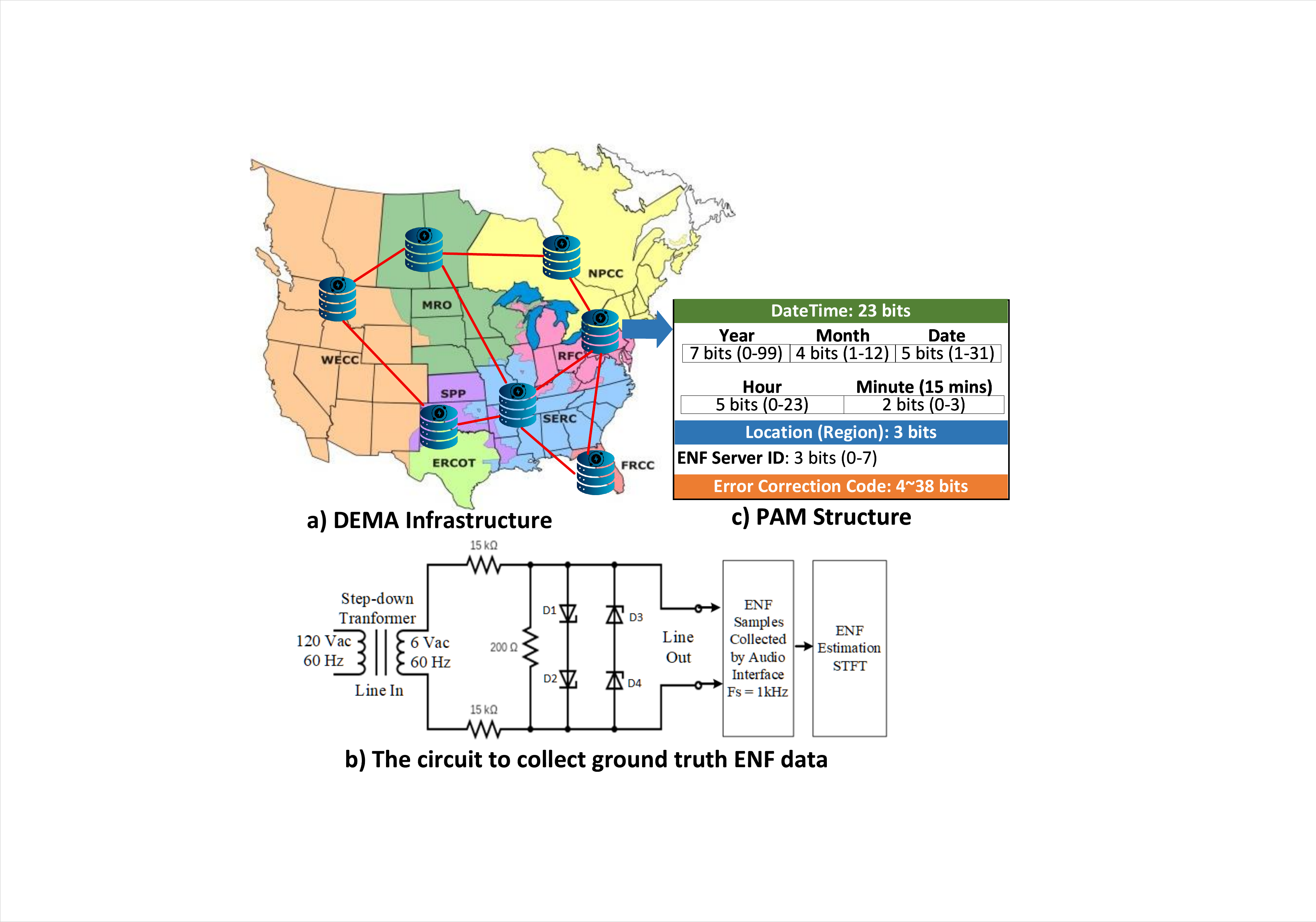}
    \caption { \label{fig:ammstruct} DEMA Infrastructure and PAM Data Structure.} 
   \end{center}
    \vspace{-10 pt}
\end{figure}

To prevent adversaries from tampering with training data or creating fake digital content for illegal purposes, we employ a physical anchor fabric that generates verifiable PAMs linked to external ground-truth references, like estimated ENF fluctuations.
Unlike traditional metadata that is client-generated and easily forged, our PAM is not a direct client-sensed fingerprint of the local ENF signal.
Instead, it is a context provenance token built from discrete parameters: time ($t$), location ($l$), and ENF Server ID ($ids$), sourced from a trusted ground-truth ENF fabric such as DEMA \cite{nagothu2023dema}.
As illustrated in Figure \ref{fig:ammstruct}-c), these parameters are combined with an error correction code (ECC) $e$ to form a binary string representing the environmental context at the time of recording.

Figure \ref{fig:ammstruct}-a) shows a DEMA infrastructure based on North American power transmission grid.
Each regional power grid relies on a dedicated server to store and manage ground truth ENF data within its network.
Figure \ref{fig:ammstruct}-b) demonstrates a specifically designed circuit that can be used for ground truth ENF data acquisition at the region server \cite{nagothu2023lightweight}.
The DEMA system allows for a regional ENF source search given time and location.
Thus, it provides physical ground truths to support data authentication in cyberspace. 
This mechanism naturally fits federated learning scenarios, where distributed clients can anchor their data provenance to geographically localized ENF references through a context provenance token.

To comply with Kerckhoffs' principle and prevent forgery, this raw PAM structure is not embedded directly. Instead, it undergoes a keyed cryptographic transformation defined as
\begin{equation}
W_{in} = \mathsf{HMAC}_{K}(\mathrm{PAM}),
\end{equation}
where $K$ denotes a client-side secret key created by the data owner and secretly shared with the physical anchor fabric, and HMAC is instantiated as HMAC-SHA256.
Consequently, the scheme establishes a context-level anchor (a soft link) to the physical infrastructure via the context-provenance token PAM. 
Crucially, the watermark itself is a cryptographic hash of the provenance claim, ensuring that even if the PAM structure is public, the keyed watermark $W_{in}$ cannot be forged without the secret key $K$.
Subsequently, the FMGAN framework embeds this cryptographically secured watermark $W_{in}$ into images.

The anchor verification pipeline begins with the watermark extraction phase, which recovers the keyed watermark $W_{in}$ from the watermarked image and forwards it, along with auxiliary metadata, to the physical anchor infrastructure.
If the extraction module fails to retrieve $W_{\text{in}}$, the pipeline immediately terminates and returns a verification failure status.
Upon successful extraction of $W_{\text{in}}$, the physical fabric parses the auxiliary metadata to extract the temporal ($t'$) and spatial ($l'$) parameters. These parameters serve as indices to query the secure physical fabric and retrieve the corresponding reference Physical Anchor Metadata ($PAM'$).
To verify that the image matches its asserted environmental context, the system evaluates the cryptographic integrity of the asset by validating the following condition:$\mathsf{HMAC}_{K}(PAM^{'}) = W_{in}$.
Upon satisfying this condition, the pipeline outputs a positive verification result, thereby certifying that 
the image carries a key-valid context-provenance token for the asserted time, location, and server, that is, that a holder of the secret key vouched for that context. It does not by itself certify that the image content is genuine; see Limitations. 
Conversely, a mismatch yields a negative verification outcome, indicating that the data has been tampered with or fabricated.

\subsection{The Watermarking Model}

Client-side training data are typically heterogeneous and visually complex, encompassing diverse textures, varying color saturation levels, and inconsistent lighting conditions, thereby leading to challenges including data heterogeneity, feature inconsistency, and domain shifts \cite{li2021fedbn}.
The complexities make embedding watermarks more difficult, particularly in maintaining both visual invisibility and robustness.

\begin{figure}[htbp]
  \hspace{-1.0 em}  \includegraphics[trim=65 130 60 70, clip, width=0.50\textwidth]{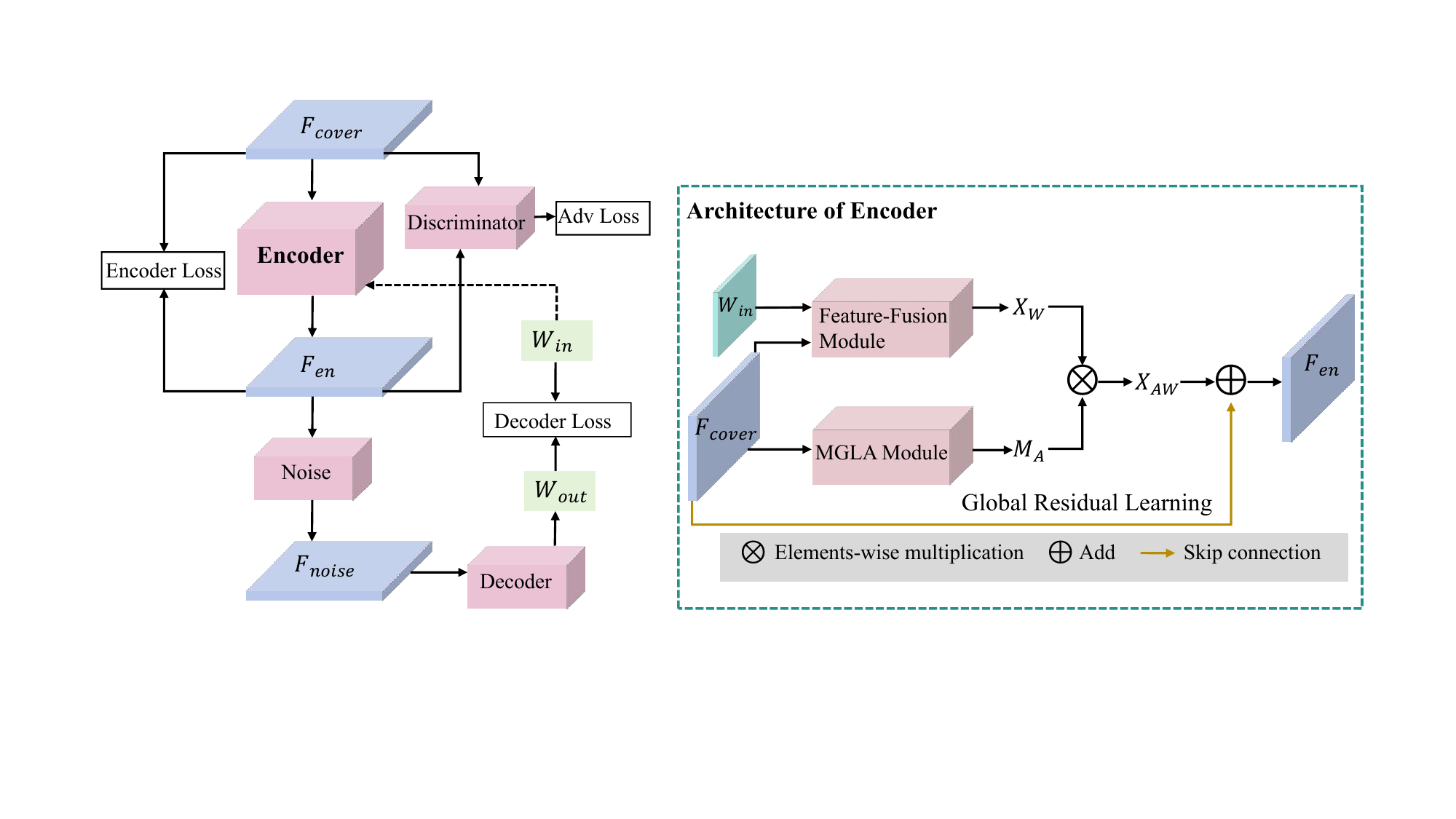}
    \caption { \label{fig:architecture} Architecture of FMGAN and its Encoder. The encoder includes (1) a Feature-Fusion Module (FFM) and (2) a Mamba-Guided Linear Attention Module (MGLA), which leverages linear attention to embed watermarks into low-saliency regions of the original image.  } 
    \vspace{-10 pt}
\end{figure}



As illustrated in Figure \ref{fig:architecture}, the FMGAN model consists of the encoder $E_{\theta}$, the decoder $D_{\beta}$, the noise layer $Noise$, and the discriminator $DC_{\delta}$. The parameters ${\theta}$, ${\beta}$, and ${\delta}$ correspond to the trainable weights of the encoder, decoder, and discriminator, respectively, and are iteratively refined to improve watermark invisibility and robustness. Given an input image $F_{cover}$ of size $H \times W \times C$ and a binary watermark $W_{in}$ of length $L$, the encoder $E_{\theta}$ produces a watermarked image $F_{en}$.

The discriminator $DC_{\delta}$ determines how closely $F_{en}$ resembles $F_{cover}$ by evaluating their similarity, guiding $E_{\theta}$ to generate a more realistic $F_{en}$. To enhance robustness, different types of distortions are applied through $Noise$. Simultaneously, adversarial training is performed, where the decoder $D_{\beta}$ extracts the watermark $W_{out}$ from the noisy image $F_{noise}$, ensuring that $W_{out}$ closely matches $W_{in}$.

\subsubsection{Encoder}





The encoder $E_{\theta}$ functions to embed the watermark into the original image $F_{cover}$ while maintaining high image quality and ensuring resilience against different types of attacks. To improve training efficiency, a residual structure is incorporated, featuring a global residual skip connection. Specifically, the encoded image is generated as $F_{en}=E(F_{cover}, W_{in})$, where $E(\cdot)$ represents the encoding process, as illustrated in Fig.~\ref{fig:architecture}.

The encoder is composed of two primary components: the Feature Fusion Module (FFM) and the Mamba-Guided Linear Attention Module (MGLA). The FFM overcomes the drawback of capturing only a limited range of image features by extracting both shallow and deep features across multiple layers. The fused feature $X_{W}$ is obtained as $X_{W}=E_{FFM}(F_{cover}, W_{in})$, where $E_{FFM}$ denotes the operation of the FFM.
The FFM consists of multiple dense blocks, each comprising Batch Normalization (BN), Rectified Linear Unit (ReLU) activation, and convolutional layers in sequence. By leveraging dense connections, the FFM facilitates feature reuse, thereby significantly enhancing the encoder's representation capability.

\begin{figure}[htbp]
  \hspace{1.0em}  \includegraphics[trim=60 100 170 50, clip, width=0.45\textwidth]{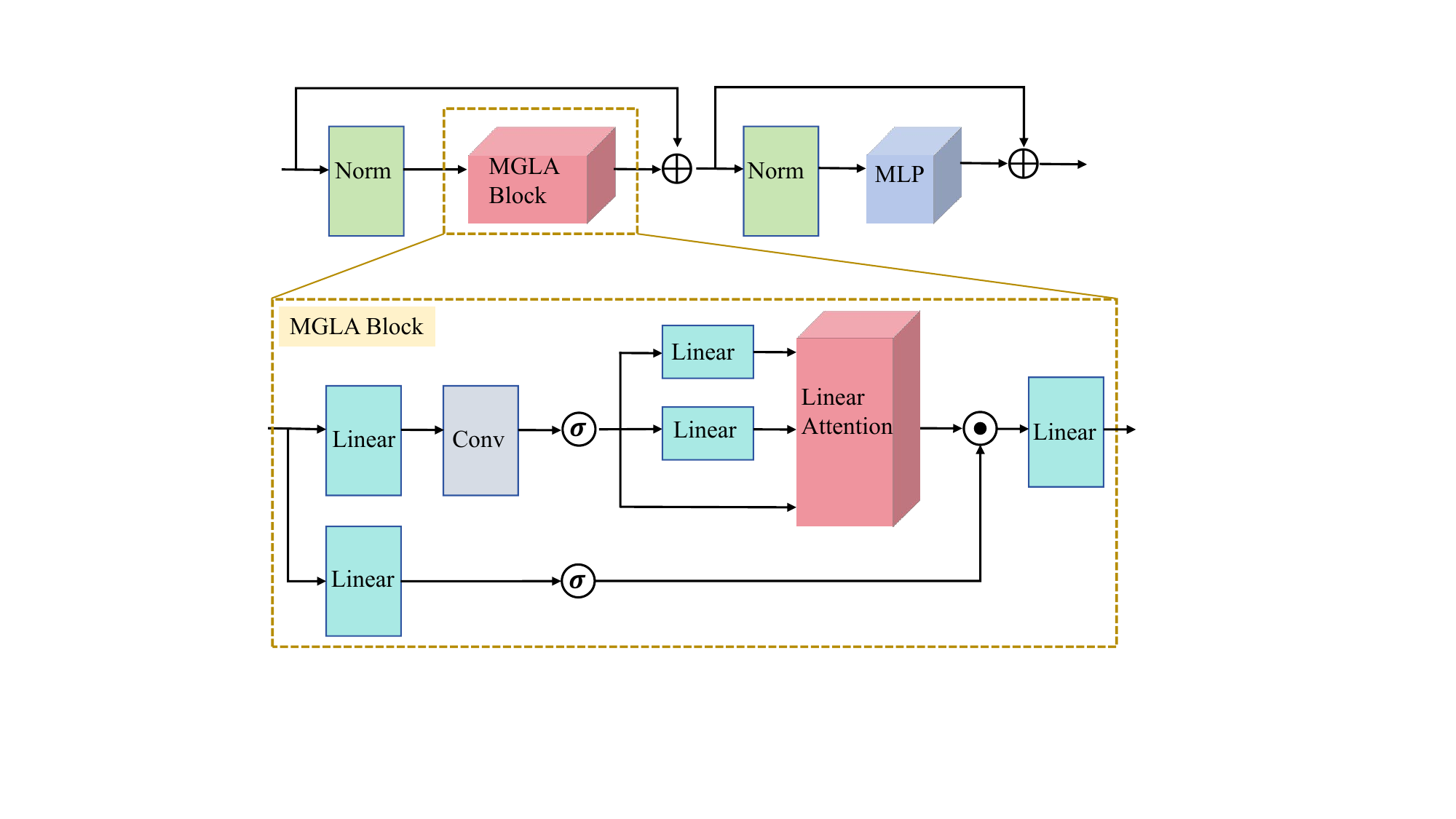}
    \caption { \label{fig:MGLA} Overview of the MGLA architectural design. } 
    \vspace{-10 pt}
\end{figure}

The way a watermark is embedded in an image can influence both its visual quality and robustness. Placing it in less noticeable areas helps maintain the image’s quality, while embedding it in textured regions strengthens its resistance to attacks. 
To address the challenge of distortion, we developed the Mamba-Guided Linear Attention Module (MGLA), which optimizes watermark placement to minimize visual impact (see Figure \ref{fig:MGLA}).
Mamba \cite{gu2024mamba} is a state-space model (SSM) for efficient sequence modeling, achieving strong performance on long-sequence tasks with reduced computational and memory complexity compared to transformers, while remaining competitive across diverse learning tasks. Prior studies indicate that its effectiveness mainly stems from the forget-gate mechanism and a modified block design. However, the recurrent nature of the forget gate introduces causal constraints, making it suboptimal for non-causal vision applications.

To overcome this limitation, our MGLA replaces the forget gate with parallelizable positional encoding schemes (e.g., LePE, CPE, or RoPE), preserving positional awareness while enabling fully parallel computation. Moreover, MGLA inherits Mamba’s block design by integrating depth-wise convolutions and gating mechanisms into a linear attention framework. This design facilitates efficient feature extraction and improved representation capacity.
By reinterpreting the selective SSM as a specialized form of linear attention, MGLA maintains the efficiency advantages of SSMs while achieving superior performance on vision tasks. Formally, the state-space model maps an input $x_i \in \mathbb{R}$ to an output $y_i \in \mathbb{R}$ via a hidden state $h_i$, and the selective SSM in Mamba can be rewritten as follows:

\begin{equation}
\begin{aligned}
    {h}_i &= \widetilde{{A}}_i \odot {h}_{i-1} + {B}_i(\Delta_i \odot x_i),  \\
    x_i, \Delta_i &\in \mathbb{R}, \ \widetilde{{A}}_i, {B}_i, {h}_{i-1}, {h}_i \in \mathbb{R}^{d \times 1}.\\
    y_i &= {C}_i {h}_i + D \odot x_i,\\
    y_i &\in \mathbb{R}, \ {C}_i \in \mathbb{R}^{1 \times d}, \ D \in \mathbb{R}.
\end{aligned}
\label{Eq:3}
\end{equation}
where $\widetilde{{A}}_i$ is the weight matrix for the previous hidden state, and ${B}_i$, ${C}_i$ are also the weight matrix; $D$ denotes the Scalar skip-connection weight; $\Delta_i$ denotes the timescale parameter; $\odot$
denotes the Hadamard product.

Mamba can be considered as a variant of linear-attention Transformers with specialized attention dynamics and a modified block design. The linear attention formulation is given by:

\begin{equation}
\begin{aligned}
    S_i &= 1 \odot S_{i-1} + K_i^\top (1 \odot V_i), \\
    y_i &= \frac{Q_i S_i}{Q_i Z_i} + 0 \odot x_i.
\end{aligned}
\label{Eq:4}
\end{equation}

Under this formulation, MGLA exhibits a strong structural correspondence with linear attention. Building upon this framework, the MGLA block further integrates depth-wise convolutions and gating mechanisms. The overall computational complexity is

\begin{equation}
\begin{aligned}
    \Omega ( \mathrm{MGLA} ) = 2 N C^{2} + 2 N C d + k^{2} N C + 8 N C^{2}
\end{aligned}
\label{Eq:5}
\end{equation}
where the terms correspond to input/output projections, linear attention, depth-wise convolution, and MLP operations, respectively. 
Consequently, MGLA achieves efficient global modeling with linear complexity $O(N)$ and fully parallelizable computation.




The MGLA module adaptively reweights feature regions based on their importance while modeling global context. The resulting attention mask dynamically modulates watermark strength across spatial locations, achieving a balanced trade-off between imperceptibility and robustness. Specifically, leveraging the global features of $F_{cover}$, the MGLA module generates an attention mask $M_{A}=E_{MGLA}(F_{cover})$ to assist in producing the encoded image.

This adaptive modulation allows $F_{en}$ to be dynamically adjusted, reducing image distortion caused by watermark embedding and minimizing visible perturbations. The attention mask $M_{A}$ is then applied to reweight the fused feature representation $X_{W}$. Finally, the encoded image is obtained through a global residual skip connection as $F_{en}=F_{cover}+X_{W}\times M_{A}$.
The encoding loss function $L_{E}$ aims to reduce the discrepancy between $F_{cover}$ and $F_{en}$ by adjusting the parameter $\theta$. It consists of two components: the image reconstruction loss and the visual loss
\begin{equation}
    L_{E}=\epsilon_{1}MSE(F_{cover}, F_{en})+\epsilon_{2}SSIM(F_{cover}, F_{en})
\label{Eq:5}
\end{equation}
where $MSE()$ represents the mean-square error function, and $SSIM()$ denotes the structural similarity index metric. The coefficients $\epsilon_{1}$ and $\epsilon_{2}$ represent the weighting factors for the image reconstruction loss and the visual loss, respectively.


\subsubsection{Decoder for Watermark Retrieval}
The noise module, denoted as $Noise$, introduces various types of differentiable noise during the iterative training process to improve watermarking robustness. 
Specifically, the noised feature map is generated as $F_{noise}=Noise(F_{en},N_{train})$, where $N_{train}$ denotes the trained noise.

Subsequently, the decoder $D_{\beta}$ is applied to extract the embedded watermark from the noised features, yielding $W_{out}=D_{\beta}(F_{noise})$.

The decoder is trained to enhance watermark extraction robustness by minimizing the discrepancy between the extracted watermark $W_{out}$ and the original watermark $W_{in}$, with respect to the decoder parameters $\beta$. The decoding loss $L_{D}$ is defined as:
\begin{equation}
L_{D}=\frac{\sqrt{(W_{in}-W_{out})^2}}{L}.
\label{Eq:DecoderLoss}
\end{equation}

Watermark extraction operates as the reverse process of embedding, utilizing a structure similar to the FFM to retrieve the watermark from the encoded image features. 
Excessive feature distortion may degrade decoding accuracy; therefore, decoder feedback is leveraged during noise-aware training to guide the encoder away from fragile feature regions. This collaborative optimization enables reliable watermark recovery without requiring access to the original image.








\subsubsection{Discriminator}

The discriminator plays a crucial role in ensuring the watermark embedded within the image is both inconspicuous and resilient. The discriminator, denoted as $DC_{\delta}$, is responsible for distinguishing between genuine (unwatermarked) image $F_{cover}$ and fake (watermarked) image $F_{en}$.
The discriminator loss assesses its ability to correctly distinguish between real and fake images, while the encoder loss motivates the encoder to generate watermarked images that become progressively harder for the discriminator to classify as fake. The discriminator adjusts its parameter ${\delta}$ to reduce the probability of incorrect classifications. The adversarial loss, $L_{A}$, is defined as: 
\begin{equation} 
L_{A}=\mathbb{E}{p \sim F{cover}} [\log(1-P(F_{en}))] \label{Eq:9} 
\end{equation} 
where $\mathbb{E}$ represents the expectation function, which computes the mean value of the sampled data $p$ over the probability distribution $F_{cover}$, and $P(\cdot)$ indicates the probability that $F_{en}$ contains $W_{in}$.
Through this adversarial training loop, the discriminator compels the encoder to fine-tune the watermark embedding, making it more resistant to perceptual attacks, such as visual removal or distortions caused by compression and noise. 
Collectively, the encoder, discriminator, and decoder contribute to the system’s robustness and practical applicability, forming the foundation for comprehensive experimental evaluation.

\subsection{LZKSA Security Protocol}
We adopt the input-verification framework of LZKSA~\cite{lu2025lzksa}, which constructs public-coin $\Sigma$-protocols directly over RLWE commitments. Instead of reducing each constraint to a generic zero-knowledge circuit, LZKSA exploits algebraic relations compatible with the response form of $\Sigma$-protocols. Consequently, the same proof flow establishes both consistency with the committed input and satisfaction of $L_\infty$-, $L_2$-, and cosine-similarity constraints (Algorithm~\ref{alg:lzksa_overview_func}).

Let $R_q$ be the underlying cyclotomic ring. A client packs its input into $\mathbf{x}\in R_t$ and computes
\begin{equation}
\mathbf{c}=\Com(\mathbf{a},\mathbf{s},\mathbf{x},\mathbf{e};q,t)
:=[\mathbf{a}\mathbf{s}+\mathbf{x}+t\mathbf{e}]_q,
\label{eq:rlwe-commit}
\end{equation}
where $\mathbf{a}\in R_q$ is public, $\mathbf{s}\in R_q$ is secret, and $\mathbf{e}$ is small. Under RLWE, $\mathbf{c}$ is computationally hiding and binding. Moreover, the ciphertext submitted for secure aggregation is directly reused as the commitment, avoiding a separate commitment layer.

\begin{algorithm}[t]
\caption{LZKSA-Based Input Verification}
\label{alg:lzksa_overview_func}
\DontPrintSemicolon
\KwIn{$\mathbf{x}$, public $(\mathbf{a},\mathbf{y})$, bounds $(B_\infty,B_2)$, threshold $\alpha$}
\KwOut{Accept or Reject}

\textbf{Client $i$:};
$\mathbf{s}\leftarrow\sum_{i<j}\mathbf{s}*{ij}-\sum*{i>j}\mathbf{s}*{ij}$;;
$\mathbf{c}\leftarrow\Com(\mathbf{a},\mathbf{s},\mathbf{x},\mathbf{e};q,t)$;;
$\pi*{\rm key}\leftarrow
(\ZKAgg(\mathbf{s}),\ZKCli(\mathbf{s},{\mathbf{s}*{ij}}*j))$;;
$\pi*\infty\leftarrow
\ZKInf(\mathbf{a},\mathbf{c};\mathbf{x},\mathbf{s},\mathbf{e},B*\infty)$;;
$\pi_2\leftarrow
\ZKTwo(\mathbf{a},\mathbf{c};\mathbf{x},\mathbf{s},\mathbf{e},B_2)$;;
$\pi_{\cos}\leftarrow
\ZKCos(\mathbf{a},\mathbf{c},\mathbf{y};
\mathbf{x},\mathbf{s},\mathbf{e},\alpha,B_2)$;;
Send $(\mathbf{c},\pi_{\rm key},\pi_\infty,\pi_2,\pi_{\cos})$ to the server.;

\textbf{Server:};
$\mathsf{acc}\leftarrow
\Verify(\mathbf{c},\mathbf{y},B_\infty,B_2,\alpha;
\pi_{\rm key},\pi_\infty,\pi_2,\pi_{\cos})$;;
\If{$\mathsf{acc}=0$}{\Return Reject}
\Return Accept and aggregate $\mathbf{c}$;
\end{algorithm}

For secure aggregation, client $i$ derives its masking key as
$\mathbf{s}_i=\sum_{i<j}\mathbf{s}_{ij}-\sum_{i>j}\mathbf{s}_{ij}$.
The protocol $\ZKAgg$ proves global cancellation,
$\sum_i\mathbf{s}_i=0$, while $\ZKCli$ proves the local decomposition of
$\mathbf{s}_i$ and the pairwise consistency
$\mathbf{s}_{ij}=\mathbf{s}_{ji}$. The latter allows an inconsistent client
to be identified, which cannot be achieved from the global cancellation
condition alone. We omit the client subscript below.

For the $L_\infty$ constraint, the prover samples a bounded blinding
element $\boldsymbol{\mu}$ and, after receiving challenge $\gamma$, returns
$\mathbf{S}=\boldsymbol{\mu}+\gamma\mathbf{x}.$
The verifier checks the corresponding ring-linear consistency equation and
$
|\mathbf{S}|_\infty\le(\gamma+1)B_\infty.
$
With the protocol-prescribed challenge range, acceptance implies
$|\mathbf{x}|_\infty\le B_\infty$: an out-of-range integral coefficient
cannot be hidden by bounded $\boldsymbol{\mu}$. Thus, no independent
per-coordinate range proofs are required.

For $|\mathbf{x}|_2\le B_2$, LZKSA proves the equivalent squared-norm
bound. Using coefficient-wise multiplication $\circ$, the verifier checks
\begin{equation}
\left|\boldsymbol{\mu}\circ\boldsymbol{\mu}
+\gamma(\mathbf{x}\circ\mathbf{x})\right|_1
\le(\gamma+1)B_2^2,
\label{eq:l2-implication}
\end{equation}
where $|\boldsymbol{\mu}|_2\le B_2$ and $\gamma>B_2^2$. Because squared
coefficient norms are integral, this inequality implies
$|\mathbf{x}|_2^2\le B_2^2$. RLWE commitments are not multiplicatively
homomorphic; therefore, LZKSA introduces commitments to the required
quadratic and cross terms, such as
$\mathbf{x}\circ\mathbf{x}$,
$2\mathbf{x}\circ\boldsymbol{\mu}$, and
$\boldsymbol{\mu}\circ\boldsymbol{\mu}$, and binds them through additional
linear consistency equations.

Finally, a cosine constraint against public reference $\mathbf{y}$ is
reduced to a bounded scalar comparison. Let
\begin{equation}
X_Y=\langle\mathbf{x},\mathbf{y}\rangle^2,
X_X=\alpha^2|\mathbf{y}|_2^2|\mathbf{x}|_2^2.
\label{eq:cosine-reduction}
\end{equation}
A threshold on squared cosine similarity is equivalent to proving the
appropriate sign of $v=X_Y-X_X$ (or $X_X-X_Y$ for the reverse inequality).
The value $|\mathbf{x}|_2^2$ is obtained from the validated
$\mathbf{x}\circ\mathbf{x}$ using a coefficient-summing operator.
Similarly, $\langle\mathbf{x},\mathbf{y}\rangle$ is derived from
$\mathbf{x}\circ\mathbf{y}$, and its square is linked through auxiliary
commitments. A dedicated RLWE range proof then establishes
$0\le v<2^p$ for a suitable $p$ by encoding the bit decomposition of $v$
as compressed inner-product relations.

Hence, LZKSA provides constraint-specialized proofs that are directly
bound to aggregation ciphertexts. It avoids generic circuit reductions and
discrete-log commitments, while preserving zero knowledge of
$\mathbf{x}$ and reducing communication and verification costs.

\subsection{Cross-Layer Security Complementarity}

The Keyed PAM watermarking layer and the LZKSA-based secure aggregation layer are designed to protect distinct yet complementary components of the federated learning pipeline. 
They jointly address orthogonal threat vectors that commonly arise in real-world federated learning deployments. \textbf{PAM watermarking mitigates risks associated with dataset leakage, data theft, and unauthorized reuse of sensitive images, while LZKSA defends against malicious gradient manipulation, poisoning at the optimization level, and aggregation-time attacks}. Together, they provide a broader security coverage than either approach alone.
The proposed framework establishes an \textbf{end-to-end protection model} that aligns with real-world FL deployment constraints and adversarial capabilities.


\section{Experimental Results}

In this section, we evaluate the proposed dual-layer framework at both component and system levels. While the watermarking and LZKSA secure aggregation scheme offer complementary guarantees, isolated evaluation is insufficient to demonstrate robustness under realistic adversarial settings. Accordingly, we conduct (1) component-level analysis, including watermark imperceptibility and robustness, as well as the correctness and efficiency of the LZKSA protocol. and (2) system-level evaluation under composite attack scenarios.

\subsection{Experimental Setup}
\label{sec:exp_setup}

\textbf{Implementation Details.}
Our framework is evaluated in a simulated FL environment. Multiple clients perform local training on private datasets and submit masked gradients; an aggregation server then runs the LZKSA protocol to perform secure aggregation (SA).
The watermarking model is implemented in PyTorch on an NVIDIA Tesla V100 GPU using COCO~\cite{COCO}, ImageNet-10~\cite{deng2009imagenet}, and FFHQ~\cite{FFHQ} datasets, with all images resized to $512 \times 512 \times 3$.
The LZKSA verification prototype is implemented in Rust and evaluated on an Intel Core i7-9700 3.00 GHz CPU with 32 GB RAM, assuming symmetric computing power across clients and the aggregator.

\textbf{Evaluation Metrics.}
Watermark imperceptibility is quantified via PSNR and SSIM over five-fold cross-validation, while extraction robustness is measured by Bit Accuracy (BA), defined as the ratio of correctly recovered bits to the ground-truth payload.
SA denotes the baseline secure aggregation process under client ($\text{Key}_{\text{client}}$) and aggregator ($\text{Key}_{\text{agg}}$) authentication protocols, while $\text{ZK}_{\infty}$, $\text{ZK}_2$, and $\text{ZK}_{\text{cos}}$ represent the zero-knowledge verification bounds for the $L_\infty$-norm, $L_2$-norm, and cosine similarity of input updates, respectively.


\subsection{Component-Level Evaluation}
\label{sec:Component-Level}

\subsubsection{Evaluation of FMGAN Watermarking}

We evaluate FMGAN watermarking for data-level protection using a keyed PAM payload, independent of downstream training dynamics or gradient behavior.

\textbf{Watermark Invisibility:}
Table~\ref{tab:comparison} benchmarks FMGAN against HiDDeN~\cite{zhu2018hidden}, MBRS~\cite{jia2021mbrs}, and ReDMark~\cite{ahmadi2020redmark} using binary payloads ($w \in \{30, 64\}$-bits) across 1{,}000 COCO images.
The framework achieves a $w$-bit security level against keyless adversaries, limiting the per-trial forgery probability to $2^{-w}$ and requiring $\mathcal{O}(2^w)$ trials. We therefore enforce $w=64$ for robust security deployment, while adopting $w=30$ only for baseline capacity and imperceptibility evaluations.
FMGAN consistently outperforms all baselines.
Specifically, it preserves optimal fidelity with an average PSNR of $41.59$~dB and an SSIM of $0.9902$ at $w=64$.
Additionally, HMAC-SHA256 generation incurs a negligible overhead of $0.01$~ms, which does not bottleneck the embedding pipeline.

\begin{figure}[htbp]
   \begin{center}
    \includegraphics[trim=433 40 200 20, clip, width=0.54\textwidth]{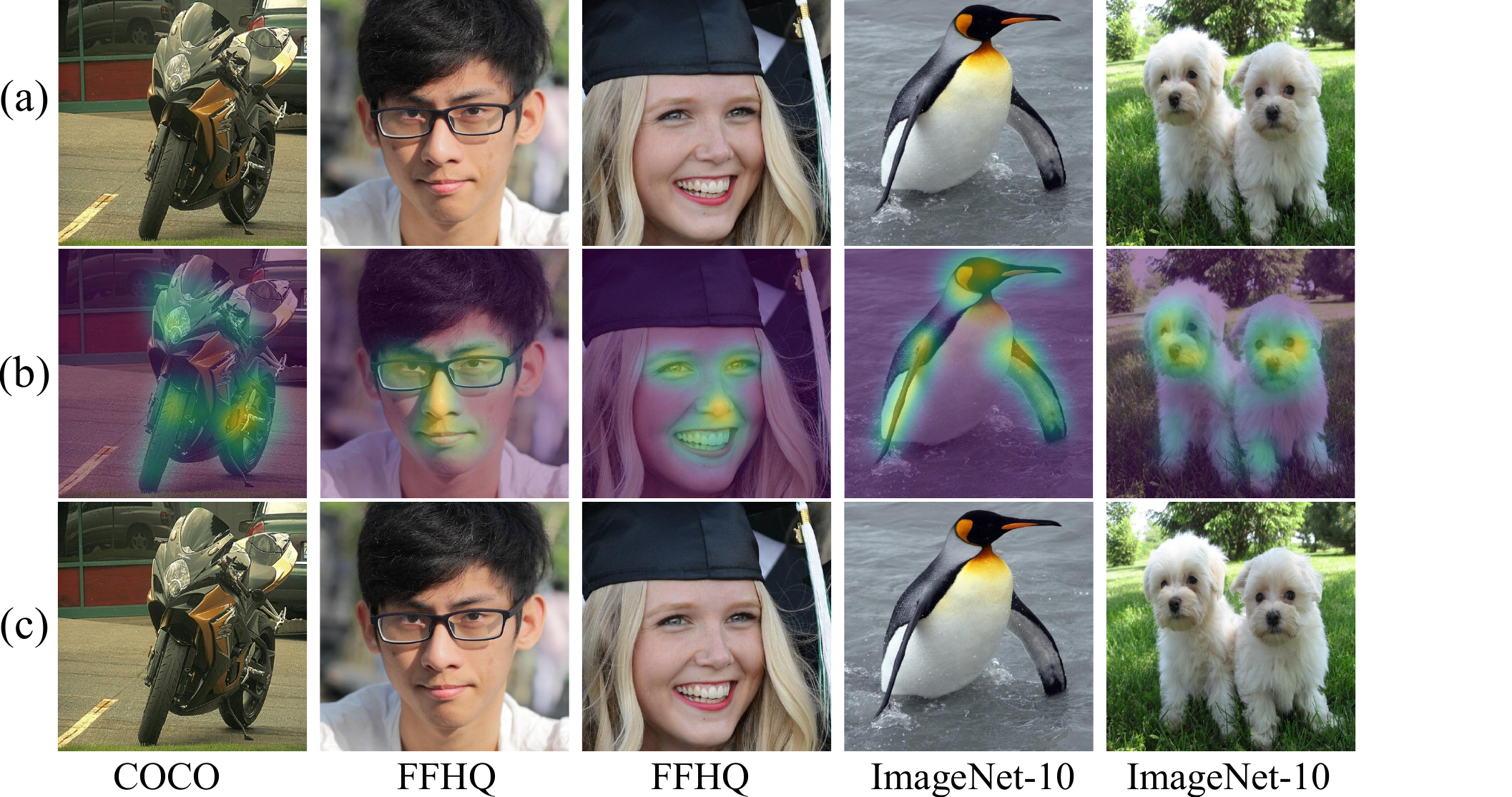}
    \caption { \label{fig:hotmap} Watermarking performance of FMGAN. (a) Original images. (b) Attention masks. (c) Watermarked images.  } 
   \end{center}
    \vspace{-10 pt}
\end{figure}

Figure~\ref{fig:hotmap} shows subjective invisibility performance.
The generated attention maps (Fig.~\ref{fig:hotmap}b) adaptively allocate stronger watermark signals to texture-rich, visually insensitive regions (indicated in yellow).
This ensures that the watermarked assets (Fig.~\ref{fig:hotmap}c) remain perceptually indistinguishable from the original images (Fig.~\ref{fig:hotmap}a).

\begin{table}[t]
\centering
\caption{Comparison of different models on the COCO dataset.}
\label{tab:comparison}
\setlength{\tabcolsep}{4pt}
\renewcommand{\arraystretch}{1.15}
\begin{tabularx}{\columnwidth}{
    p{2.2cm}
    X
    X
    X
    X
}
\toprule
\textbf{Model} &
\multicolumn{2}{c}{\textbf{$w=30$}} &
\multicolumn{2}{c}{\textbf{$w=64$}} \\
\cmidrule(lr){2-3}
\cmidrule(lr){4-5}
 & \textbf{PSNR} & \textbf{SSIM} & \textbf{PSNR} & \textbf{SSIM} \\
\midrule
HiDDeN~\cite{zhu2018hidden} & 32.03 & 0.9143 & 31.93 & 0.9137 \\
MBRS~\cite{jia2021mbrs}    & 33.29 & 0.9022 & 33.03 & 0.9025 \\
ReDMark~\cite{ahmadi2020redmark} & 37.64 & 0.9681 & 37.26 & 0.9678 \\
\textbf{FMGAN}             & \textbf{41.75} & \textbf{0.9903} & \textbf{41.59} & \textbf{0.9902} \\
\bottomrule
\end{tabularx}
\vspace{-10 pt}
\end{table}

\textbf{Watermark Robustness:}

To evaluate watermark extraction resilience to typical transmission and storage distortions, we measure Bit Accuracy (BA) across five categories: noise, compression, color, filtering, and geometric attacks.
As shown in Table \ref{tab:robustness}, FMGAN consistently outperforms HiDDeN under severe degradation vectors.
Under additive Gaussian noise ($\sigma^2 = 0.10$), FMGAN maintains a high BA of $96.37\%$, whereas HiDDeN drops to $76.55\%$. Similarly, under aggressive JPEG compression ($\text{QF}=10$), FMGAN's accuracy exceeds HiDDeN's by $20.77\%$.
Our model also demonstrates superior robustness against both color distortions (salt-and-pepper noise, brightness/contrast adjustments, histogram equalization) and geometric attacks (cropping, scaling, rotation, shearing).
Furthermore, this robust extraction performance remains highly stable when scaling the payload to $64$~bits, confirming an optimized trade-off between capacity, imperceptibility, and robustness.



\begin{table}[t]
\centering
\caption{Robustness evaluation of the proposed FMGAN model under various attacks.}
\label{tab:robustness}
\renewcommand{\arraystretch}{1.15}
\begin{tabularx}{\columnwidth}{
    p{1.7cm}
    p{1.7cm}
    p{1.7cm}
    X
    X
}
\toprule
\textbf{Attack} & \textbf{Parameter} &
\textbf{HiDDeN (\%)} &
\multicolumn{2}{c}{\textbf{FMGAN (\%)}} \\
\cmidrule(lr){4-5}
 &  &  & \textbf{$w=30$} & \textbf{$w=64$} \\
\midrule
\multirow{3}{*}{Gaussian Noise}
 & 0.001 & 83.14 & 100.00 & 100.00 \\
 & 0.05  & 80.36 & 98.19 & 99.39 \\
 & 0.10  & 76.55 & 96.37 & 97.95 \\
\midrule
\multirow{3}{*}{JPEG}
 & QF=10 & 73.69 & 94.46 & 96.79 \\
 & QF=50 & 76.72 & 98.55 & 98.04 \\
 & QF=90 & 92.05 & 99.21 & 99.32 \\
\midrule
Cropping & [20,20,420,420] & 79.24 & 98.36 & 99.05 \\
\midrule
Dropout & 0.3 & 86.87 & 99.15 & 99.34 \\
\midrule
\multirow{3}{*}{Salt \& Pepper}
 & 0.001 & 89.93 & 99.24 & 99.47 \\
 & 0.05  & 82.47 & 98.26 & 98.14 \\
 & 0.10  & 79.23 & 97.59 & 98.78 \\
\midrule
\multirow{2}{*}{Rotation}
 & 45$^\circ$ & 78.56 & 97.95 & 98.65 \\
 & 90$^\circ$ & 75.67 & 96.77 & 98.19 \\
\midrule
\multirow{2}{*}{Median Filter}
 & [2,2] & 88.22 & 98.80 & 99.03 \\
 & [3,3] & 80.74 & 97.41 & 98.37 \\
\midrule
\multirow{2}{*}{Brightness}
 & 1.1 & 92.19 & 98.94 & 99.02 \\
 & 1.3 & 83.46 & 97.28 & 98.79 \\
\midrule
\multirow{2}{*}{Contrast}
 & 1.0 & 84.64 & 97.08 & 97.82 \\
 & 2.0 & 77.53 & 96.66 & 97.23 \\
\midrule
\multirow{2}{*}{Scaling}
 & 0.5 & 76.39 & 98.54 & 99.07 \\
 & 2.0 & 81.45 & 96.80 & 96.66 \\
\midrule
Shearing & [0.4,0.4] & 80.63 & 98.69 & 99.04 \\
\midrule
Histogram Eq. & 1.0 & 86.16 & 98.91 & 99.18 \\
\bottomrule
\end{tabularx}
\vspace{-10 pt}
\end{table}

\begin{table}[t]
\centering
\caption{Comparison of different models on the COCO dataset.}
\label{tab:coco_comparison_single_noavg}
\footnotesize
\setlength{\tabcolsep}{3pt}
\renewcommand{\arraystretch}{1.15}
\begin{tabular}{
p{1.45cm}
c c
c c c c c
}
\toprule
 & \multicolumn{2}{c}{\textbf{Invisibility}} 
 & \multicolumn{5}{c}{\textbf{Robustness (\%)}} \\
\cmidrule(lr){2-3}
\cmidrule(lr){4-8}
\textbf{Model}
 & \textbf{PSNR} & \textbf{SSIM}
 & \textbf{JPEG} & \textbf{Cropout} & \textbf{Dropout}
 & \textbf{Crop} & \textbf{Gauss.} \\
 & \textbf{(dB)} & 
 & \textbf{(50)} & \textbf{(30\%)} & \textbf{(30\%)}
 & \textbf{(3.5\%)} & \textbf{(2.0)} \\
\midrule
\textit{ReDMark}~\cite{ahmadi2020redmark}
 & 38.94 & 0.9673
 & 74.62 & 92.57 & 91.98 & \textbf{100.00} & 50.23 \\

\textit{DA}~\cite{luo2020distortion}
 & 33.75 & 0.8702
 & 81.75 & 91.01 & 97.93 & 93.56 & 60.08 \\

\textit{TSDL}~\cite{liu2019novel}
 & 33.53 & 0.8931
 & 76.24 & 96.59 & 97.41 & 89.04 & 98.61 \\

\textit{MBRS}~\cite{jia2021mbrs}
 & 35.79 & 0.8896
 & 91.94 & \textbf{99.98} & 99.96 & 92.69 & \textbf{100.00} \\

\textit{SSLW}~\cite{fernandez2022watermarking}
 & 34.02 & 0.8737
 & 83.03 & 79.64 & 88.17 & 50.72 & 98.95 \\

\textbf{FMGAN}
 & \textbf{41.75} & \textbf{0.9903}
 & \textbf{98.55} & 99.05 & \textbf{100.00} & 99.05 & 99.98 \\
\bottomrule
\end{tabular}
\vspace{-10 pt}
\end{table}

\begin{table}[t]
\centering
\caption{Comparison of different models on the ImageNet-10 dataset.}
\label{tab:ImageNet_comparison_single_noavg}
\footnotesize
\setlength{\tabcolsep}{2.5pt}
\renewcommand{\arraystretch}{1.15}
\begin{tabular}{
p{1.45cm}
c c
c c c c c
}
\toprule
 & \multicolumn{2}{c}{\textbf{Invisibility}} 
 & \multicolumn{5}{c}{\textbf{Robustness (\%)}} \\
\cmidrule(lr){2-3}
\cmidrule(lr){4-8}
\textbf{Model}
 & \textbf{PSNR} & \textbf{SSIM}
 & \textbf{JPEG} & \textbf{Cropout} & \textbf{Dropout}
 & \textbf{Crop} & \textbf{Gauss.} \\
 & \textbf{(dB)} & 
 & \textbf{(50)} & \textbf{(30\%)} & \textbf{(30\%)}
 & \textbf{(3.5\%)} & \textbf{(2.0)} \\
\midrule
\textit{HiDDeN}~\cite{zhu2018hidden}
 & 33.28 & 0.9105
 & 76.42 & 95.87 & 95.43 & 96.04 & 88.69 \\

\textit{DA}~\cite{luo2020distortion}
 & 34.08 & 0.8721
 & 81.56 & 91.06 & 97.85 & 93.47 & 60.02 \\

\textit{TSDL}~\cite{liu2019novel}
 & 33.42 & 0.8906
 & 76.63 & 96.41 & 97.28 & 89.11 & 98.57 \\

\textit{MBRS}~\cite{jia2021mbrs}
 & 37.06 & 0.8914
 & 91.85 & \textbf{99.96} & 99.92 & 92.57 & 99.15 \\

\textit{SSLW}~\cite{fernandez2022watermarking}
 & 35.14 & 0.8829
 & 83.15 & 79.86 & 88.73 & 61.83 & 98.71 \\

\textbf{FMGAN}
 & \textbf{41.83} & \textbf{0.9901}
 & \textbf{98.34} & 99.02 & \textbf{100.00} & \textbf{99.01} & \textbf{100.00} \\
\bottomrule
\end{tabular}
\vspace{-10 pt}
\end{table}

\textbf{Comparison With SOTA Watermarking Models:}
To further validate the effectiveness of the proposed approach, we conduct comparative experiments against several SOTA watermarking methods that incorporate noise-aware training strategies. Table \ref{tab:coco_comparison_single_noavg} presents a comprehensive comparison of invisibility and robustness on the COCO dataset. The results demonstrate that FMGAN consistently outperforms all competing SOTA methods. In particular, FMGAN achieves superior image quality while exhibiting stronger robustness than DA, SSLW, and TSDL.
Cross-dataset generalization is further examined by conducting evaluations on ImageNet-10, with comparisons against HiDDeN, MBRS, SSLW, and FMGAN summarized in Table \ref{tab:ImageNet_comparison_single_noavg}.

\begin{figure}[htbp]
   \begin{center}
    \includegraphics[trim=43 113 60 110, clip, width=0.50\textwidth]{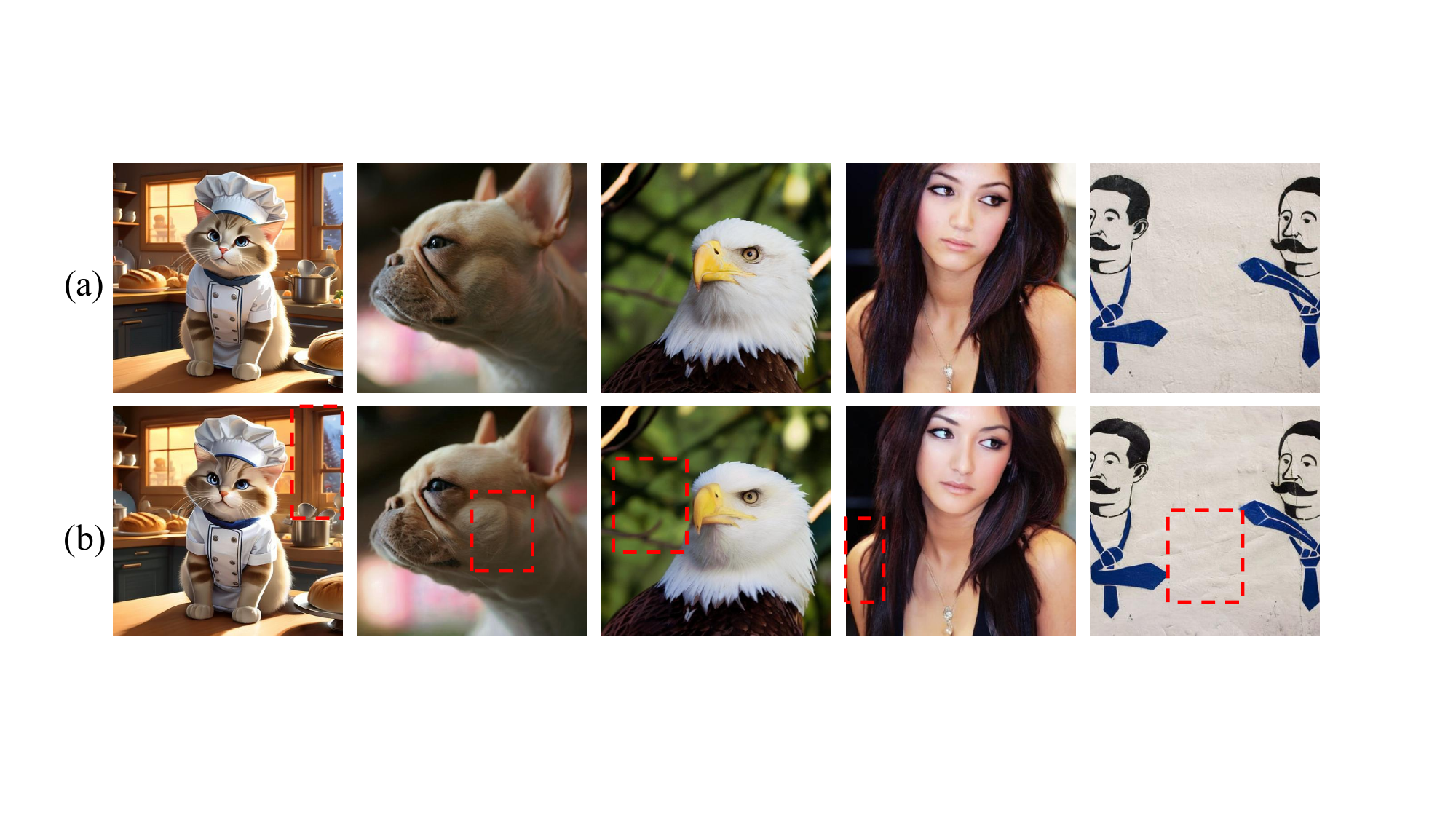}
    \caption { \label{fig:WM_Remove} Watermark Removal Attack. (a) Watermarked images. (b) Watermark-removed images. The red dashed boxes highlight visually noticeable differences. } 
   \end{center}
    \vspace{-10 pt}
\end{figure}

\textbf{Watermark Removal Attack:} We employed a state-of-the-art watermark removal method from ICLR 2025 \cite{liu2025image}, which reduced the TPR@1\%FPR metric from 1.00 to 0.02, while the Bit Accuracy also dropped significantly, from 1.00 to 0.43 on average.
Although the watermark removal attack was effective in reducing the watermark detection performance, the diffusion-based image regeneration process \cite{liu2025image} introduced perceptible visual inconsistencies.  As shown in Figure \ref{fig:WM_Remove}, while the regenerated images preserved the primary object structures and overall contours with high fidelity, noticeable discrepancies in fine-grained details remained in the background regions. These artifacts can still be identified by human observers, making the regenerated images distinguishable from the original watermarked images. Therefore, we argue that current watermark removal attacks cannot yet be considered fully successful, but instead provide valuable insights for the development of more robust future defense mechanisms.

\subsubsection{Evaluation of Secure Aggregation}
\label{sec:exp_lzksa}

To evaluate the performance of the proposed LZKSA protocol in FL scenarios, we conducted experiments on multiple widely used datasets with standard model architectures. Specifically, we trained a convolutional neural network in the Federated-MNIST dataset~\cite{caldas2018leaf}, employed LeNet-5~\cite{lecun2002gradient} and ResNet-20 convolutional neural networks in the CIFAR-10 S and CIFAR-10 L datasets~\cite{krizhevsky2009learning}, respectively, and trained an LSTM model~\cite{hochreiter1997long} in the Shakespeare text-based dataset. For the ImageNet-10 dataset, we adopted ResNet-18 \cite{he2016deep} as the backbone model.
All experiments were conducted in the same FL setting that contains clients and an aggregation server.

\begin{figure}[htbp]
  \hspace{-1.0 em}  \includegraphics[trim=40 130 60 30, clip, width=0.50\textwidth]{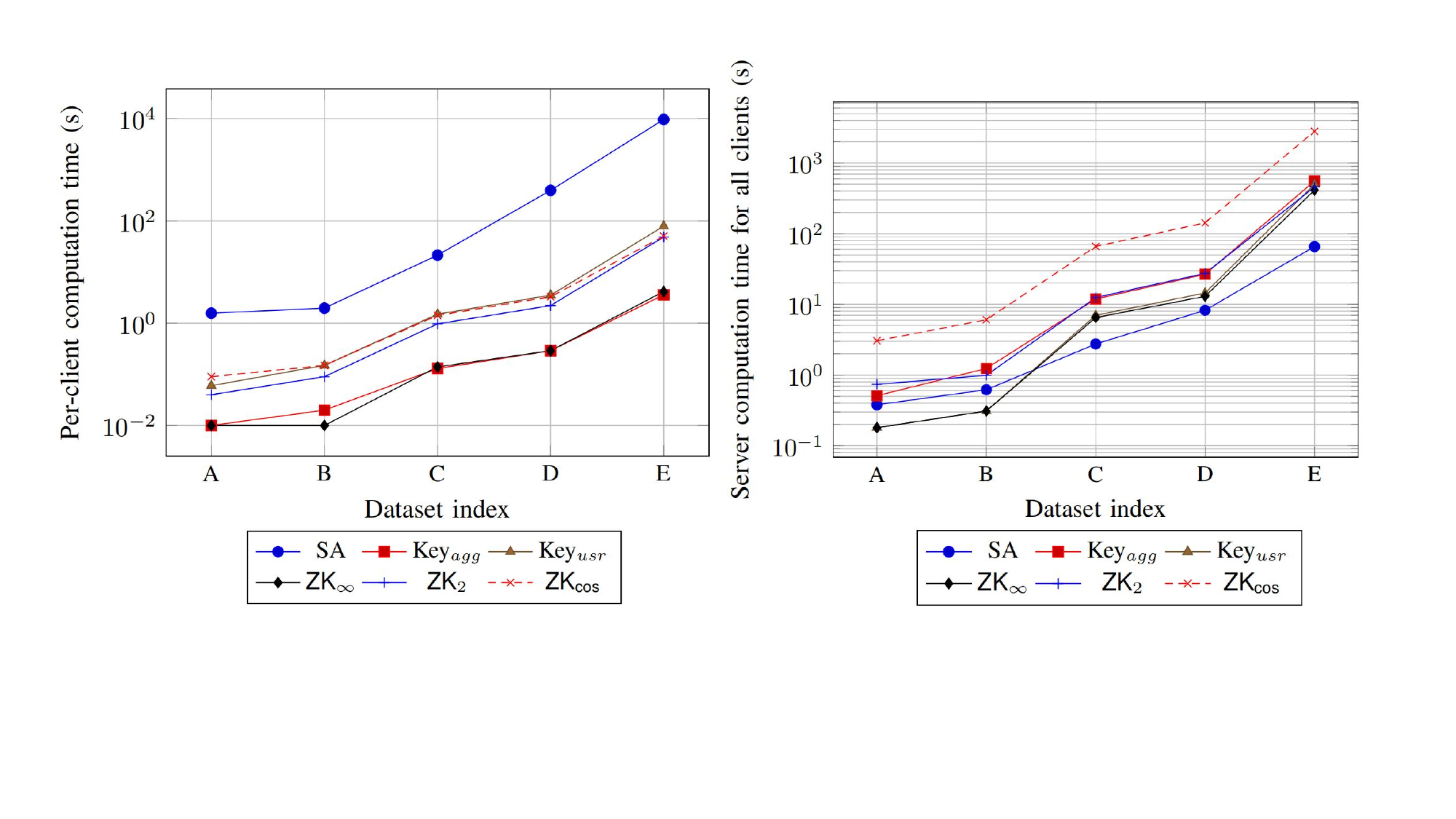}
    \caption { \label{fig:LZKSA_exp} Left: Per-client computation time comparison of the LZKSA protocols and SA.
Right: Server-side computation time across different datasets.
Both plots are obtained with 50 clients. A–E denote MNIST (19k), CIFAR-10 S (62k), CIFAR-10 L (273k), Shakespeare (818k), and ImageNet-10 (11181k), respectively.
} 
\end{figure}

Figure \ref{fig:LZKSA_exp} compares the execution time of our protocol with the standard SA scheme under different data dimensions, corresponding to MNIST, CIFAR-10 S, CIFAR-10 L, Shakespeare, and ImageNet-10, where the parameter sizes range from 19k to 11181k. We report both the per-client computation time and the server-side computation time for all clients, assuming a total of 50 participating clients.

On the client side, the experimental results in the left plot of Fig. \ref{fig:LZKSA_exp} indicate that the additional overhead introduced by the proposed verification protocols remains negligible across all data dimensions. In particular, for small- and medium-scale datasets (19k and 62k), the computation time for \textit{Key$_{agg}$}, \textit{Key$_{usr}$} and data validation protocols ($\ZKInf$, $\ZKTwo$ and $\ZKCos$) is consistently below 0.2 seconds, which is insignificant compared to the overall execution time of SA. Even for large-scale settings such as Shakespeare (818k) and ImageNet-10 (11181k), the client-side overhead increases smoothly and remains practical, which demonstrates good scalability with respect to data dimensionality.

On the server side, the computation time also grows linearly with user size, as shown in the right plot of Fig. \ref{fig:LZKSA_exp}, which is consistent with the theoretical complexity analysis. Among all protocols, $\ZKCos$ incurs the largest overhead, as it requires verification over all clients and involves more complex arithmetic operations. Nevertheless, even in the largest configuration (11181k), the server-side verification time remains within an acceptable range. Compared to the baseline SA protocol, the additional verification cost introduced by our method does not dominate the overall aggregation process.
Overall, the results in Fig. \ref{fig:LZKSA_exp} 
demonstrate that the proposed protocol achieves efficient and scalable input verification with minimal computational overhead for both clients and the server. Similar to the observations reported in LZKSA, the experimental results confirm that the added security guarantees can be obtained without sacrificing the practicality of federated learning, even under large-scale and high-dimensional settings.

\subsection{System-Level Threat Evaluation}

Real-world FL systems face composite adversarial behaviors that simultaneously target multiple layers.
We conduct a system-level threat evaluation under realistic cross-layer attack scenarios. 
We consider three representative composite attacks that jointly exploit data manipulation, gradient tampering, and cryptographic protocol abuse.

\subsubsection{Attack I: Dataset Replacement and Gradient Scaling Attack}

\begin{table}[t]
\centering
\caption{ Detection Performance under Dataset Replacement and Gradient Scaling (Attack I)}
\label{tab:attack_1}
\setlength{\tabcolsep}{2.5pt}
\renewcommand{\arraystretch}{1.15}

\begin{tabularx}
{\columnwidth}{
    p{2.1cm}
    X
    X
    X
    X
}
\toprule
\multicolumn{5}{c}{\textbf{Dataset Replacement (10\% Replaced)}} \\
\midrule
\textbf{Metric} & \textbf{Replaced Images } & \textbf{Watermarked Images }  &  &  \textbf{Overall}  \\
\midrule
Number of Images & 100 & 900 &  &  1000 \\
Detection Rate & 100\% & -- &  &  100\%  \\
FPR & -- & 0\% &  &  0\%  \\
\midrule
\multicolumn{5}{c}{\textbf{Gradient Scaling}} \\
\midrule
\textbf{Metric} & $B=0.5$ & $B=1.0$ & $B=1.5$ & $B=5.0$   \\
\midrule
\textbf{Accuracy} & 82\%-87\% & 93\%-94\% & 95\%-96\% & 33\%-34\%   \\
\bottomrule
\end{tabularx}
\end{table}

\begin{table}[t]
\centering
\caption{System Performance under Attack I}
\label{tab:attack_system}
\setlength{\tabcolsep}{4pt}
\renewcommand{\arraystretch}{1.15}

\begin{tabularx}
{\columnwidth}{
    p{2.1cm}
    X
    X
    X
    X
}
\toprule
 & No defense & FMGAN only & LZKSA only & Dual-layer framework   \\
\midrule
\textbf{Unauthorized samples detected} & 0\% & 100\% & 0\% & 100\%   \\
\midrule
\textbf{Malicious updates accepted} & 100\% & 100\% & 4\%-5\% & 4\%-5\%   \\
\bottomrule
\end{tabularx}
\vspace{-10 pt}
\end{table}

We assume that a malicious adversary compromises a client and gains full control over its local training process and gradient transmission.
The adversary replaces the local training dataset with unauthorized or irrelevant data assets and simultaneously substitutes the uploaded gradient with an abnormally large-magnitude vector to disrupt aggregation (input-out-of-range attack).

To detect dataset substitution, we simulate a $10\%$ random replacement across $1{,}000$ watermarked COCO images.
As Table~\ref{tab:attack_1} shows, our verification pipeline detects all substituted assets with $100\%$ accuracy and zero false positives.
Concurrently, gradient scaling attacks are mitigated by validating update norms against $L_{\infty}$ and $L_2$ verification bounds ($\text{ZK}_{\infty}, \text{ZK}_2$). Because bounding thresholds dictate the false-positive/false-negative trade-off, we evaluate median-derived scaling limits $B \in \{0.5, 1.0, 1.5, 5.0\}$ (Table~\ref{tab:attack_1}).
Our results indicate that $B=1.5$ yields an optimal security-utility trade-off, aligning with the empirical results of ROFL~\cite{lycklama2023rofl}.
Finally, Table~\ref{tab:attack_system} presents the system security performance under this joint threat model. The results confirm that our integrated FMGAN + LZKSA framework robustly mitigates these concurrent cross-layer attacks, significantly outperforming both standalone defenses and undefended baselines.

\subsubsection{Attack II: Forged Data Injection and Directional Gradient Manipulation}
We assume that a malicious adversary compromises a client and gains control over its local training process and gradient transmission.
The adversary injects poisoned or forged samples (e.g., backdoor triggers) into the local dataset, and simultaneously performs directional manipulation of gradients, such as sign-flipping or targeted direction amplification, to strengthen the poisoning effect during aggregation.

To detect forged data injection, we introduce $10\%$ unwatermarked images into a dataset of $1{,}000$ watermarked COCO assets.
As shown in Table~\ref{tab:attack_2}, the proposed mechanism achieves a $100\%$ detection rate with zero false positives on genuine images.
Additionally, gradient directional anomalies are constrained using cosine similarity: $\cos(\mathbf{x}, \mathbf{y}) < \alpha$, where $\mathbf{x}$ is the input gradient, $\mathbf{y}$ is the median benchmark, and $\alpha$ represents the threshold.
While setting $\alpha \to 1$ tightens security with the lower possibility of non-compliant data entering, but it at the expense of higher false-rejection rates.
The empirical evaluation identifies $\alpha = 0.8$ as the optimal threshold for balancing update utility and anomaly filtering (Table~\ref{tab:attack_2}).

\begin{table}[t]
\centering
\caption{ Detection Performance under Forged Data Injection and Directional Gradient Manipulation (Attack II)}
\label{tab:attack_2}
\setlength{\tabcolsep}{4pt}
\renewcommand{\arraystretch}{1.15}

\begin{tabularx}
{\columnwidth}{
    p{2.1cm}
    X
    X
    X
    X
}
\toprule
\multicolumn{5}{c}{\textbf{Forged Data Injection  (10\% Injected)}} \\
\midrule
\textbf{Metric} & \textbf{Injected Images } & \textbf{Watermarked Images }  &  &  \textbf{Overall}  \\
\midrule
Number of Images & 100 & 1000 &  &  1100 \\
Detection Rate & 100\% & -- &  &  100\%  \\
FPR & -- & 0\% &  &  0\%  \\

\midrule
\multicolumn{5}{c}{\textbf{Directional Gradient Manipulation}} \\
\midrule
\textbf{Metric} & $\alpha= 0$ & $\alpha= 0.5$ & $\alpha= 0.8$ & $\alpha= 1.0$\\
\midrule
\textbf{Accuracy} & 63\% & 91\% & 95\% & 86\%  \\

\bottomrule
\end{tabularx}
\vspace{-10 pt}
\end{table}


\subsubsection{Attack III: Data Noise Perturbation and Secure Aggregation Protocol Abuse}

\begin{figure*}[t]
\centering
\setlength{\subfigcapskip}{-3mm}
\subfigure[]{%
  \includegraphics[width=0.32\textwidth]{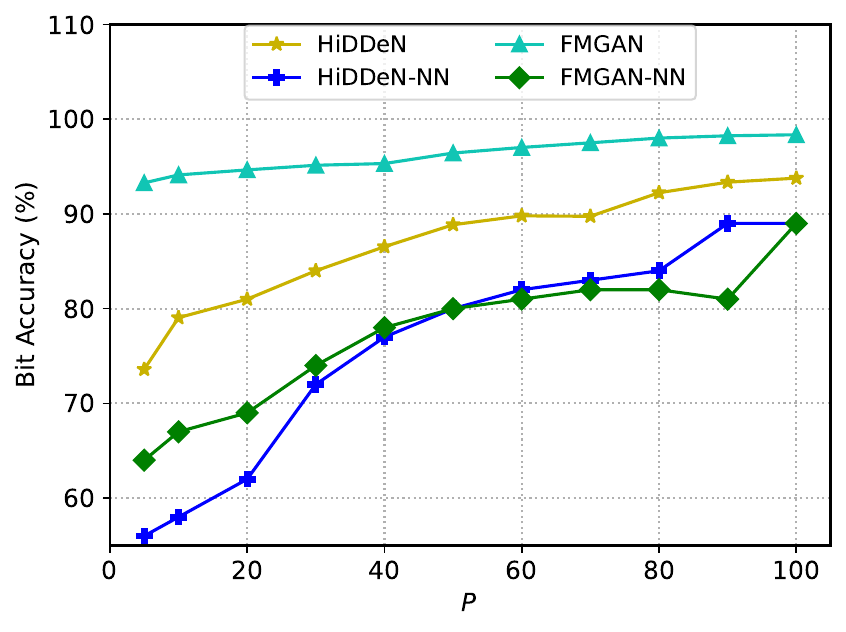}
}
\hfill
\subfigure[]{%
  \includegraphics[width=0.32\textwidth]{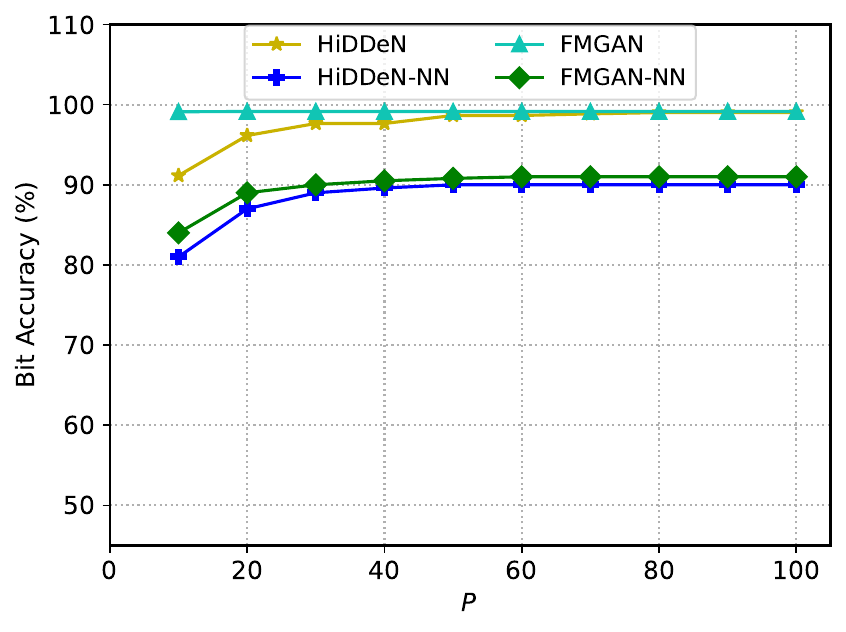}
}
\hfill
\subfigure[]{%
  \includegraphics[width=0.32\textwidth]{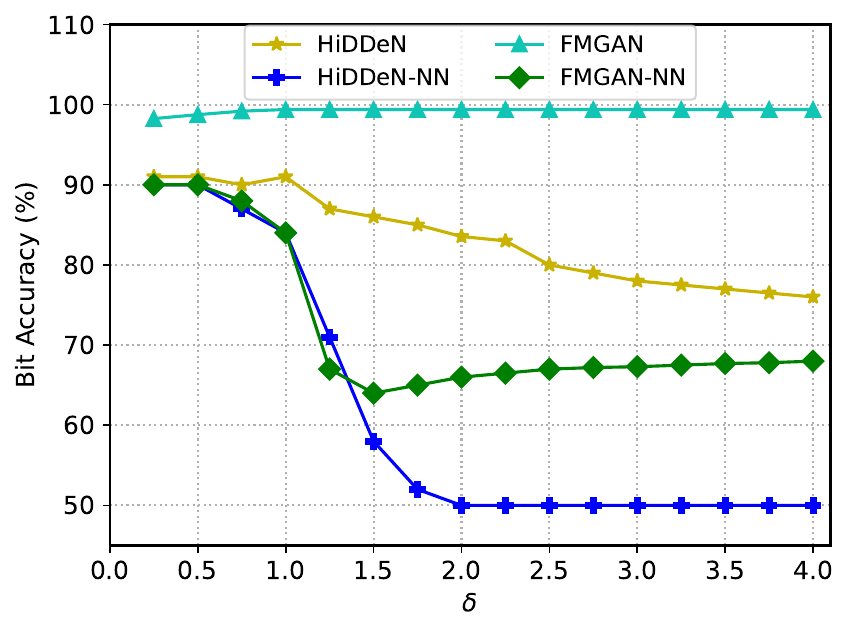}
}

\vspace{-3mm}

\subfigure[]{%
  \includegraphics[width=0.32\textwidth]{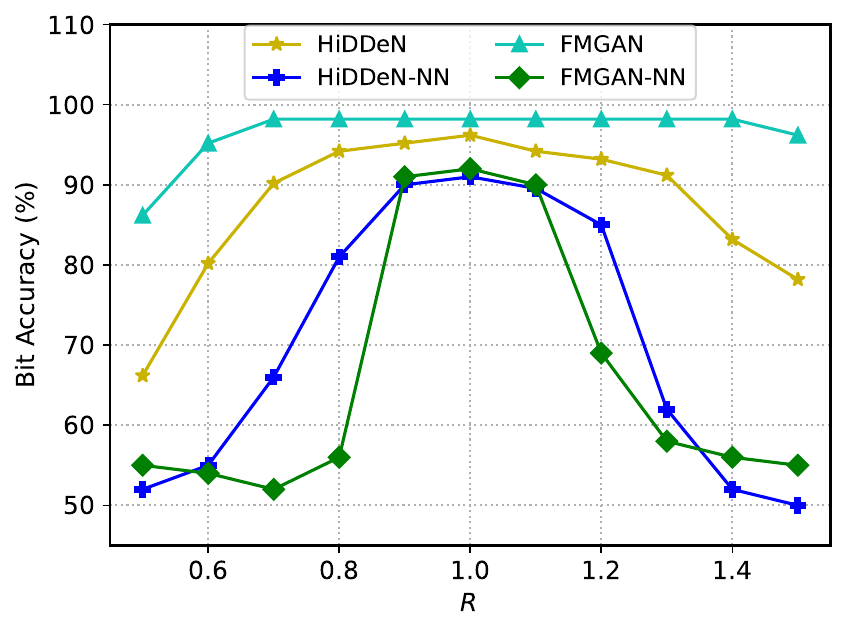}
}
\hfill
\subfigure[]{%
  \includegraphics[width=0.32\textwidth]{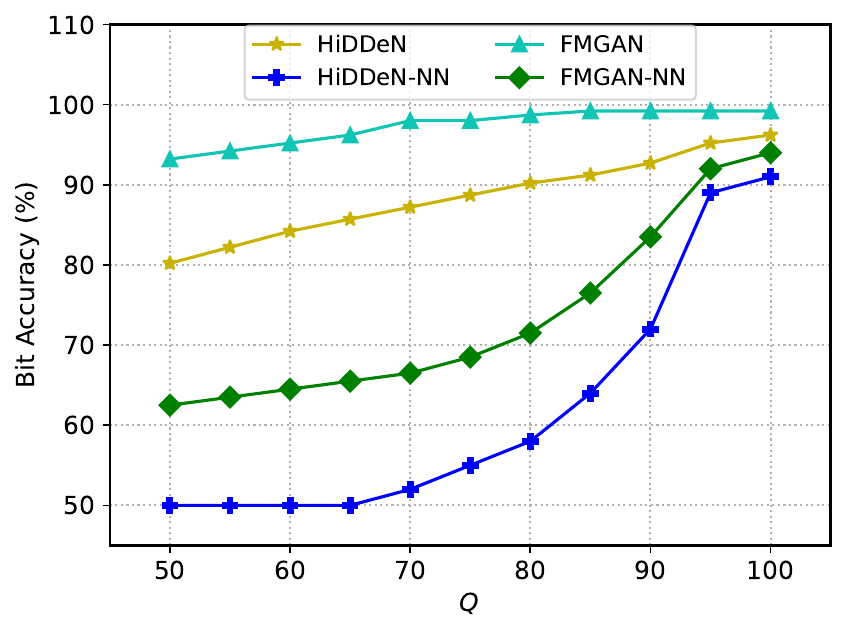}
}
\hfill
\subfigure[]{%
  \includegraphics[width=0.32\textwidth]{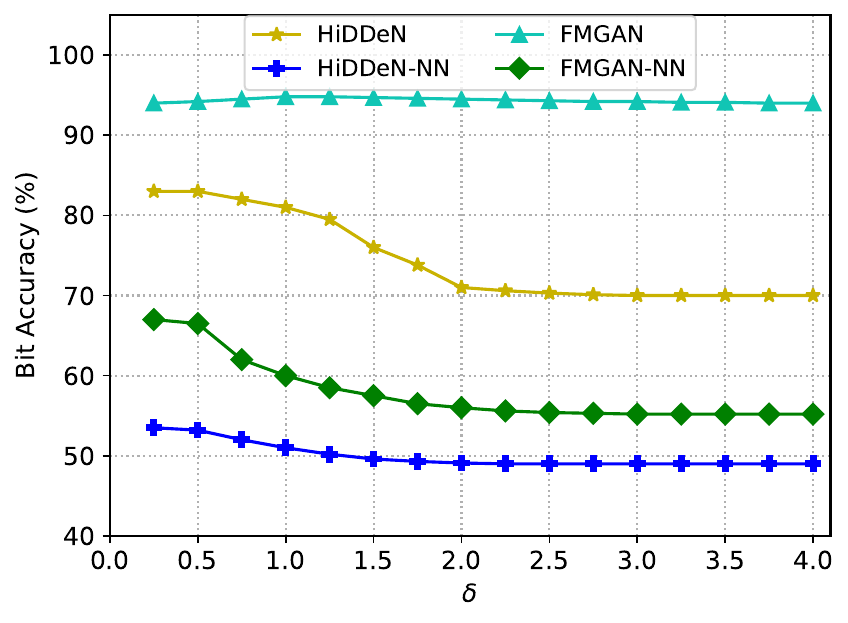}
}

\caption{Robustness evaluation under multiple intensity levels of trained noises.
(a) Crop,
(b) Dropout,
(c) Gaussian blur,
(d) Resize,
(e) JPEG compression,
(f) JPEG (Q=50) + Gaussian blur.}
\label{fig:robustness}
\vspace{-10 pt}
\end{figure*}

We assume a compromised client that targets both watermark reliability and aggregation integrity.
To degrade the detectability of watermarks, an adversary perturbs local training assets with adversarial noise. 
At the model level, it attempts to subvert the secure aggregation protocol by: 1) submitting malformed encrypted gradients, 2) executing partial aggregation replay attacks, or 3) generating structurally valid but mathematically incorrect zero-knowledge proofs.

Such aggregation-targeted attacks will be directly rejected by the server (Algorithm~\ref{alg:lzksa_overview_func}) because they would result in a mismatch between the left and right sides of the corresponding verification equation.
The probability of successfully passing verification under such attacks is negligible in the security parameter under the RLWE assumption.


We evaluate Attack III's impact on watermark detectability (Fig.~\ref{fig:robustness}) across FMGAN, HiDDeN, and their "NN" (no-noise-subnetwork) variants.
The evaluation spans varying distortion parameters: crop/dropout probability $P$, JPEG quality factor $Q$, resizing ratio $R$, and Gaussian blur variance $\delta$.
Overall, FMGAN consistently achieves the highest BA across all perturbation types and intensities, showing strong resistance to adversarial watermark suppression. This robustness advantage is particularly pronounced under representative degradation operations—namely Gaussian blur, image resizing, and JPEG compression—which are commonly exploited in practice to obscure embedded signals.

Under cropping and dropout attacks (Figs.~\ref{fig:robustness}a,b), FMGAN and FMGAN-NN consistently outperform their HiDDeN counterparts, highlighting superior resilience against content removal.
Notably, FMGAN-NN's margin over HiDDeN-NN suggests that our feature fusion mechanism preserves intrinsic structural robustness even without noise-aware training.
For Gaussian blur (Fig.~\ref{fig:robustness}c), FMGAN maintains high BA against low-pass filtering effects where baselines (particularly HiDDeN-NN) degrade rapidly; a similar robust trend holds for spatial resizing (Fig.~\ref{fig:robustness}d).
Under JPEG compression (Fig.~\ref{fig:robustness}e), FMGAN exhibits strong resilience to aggressive quantization artifacts at low quality factors.
Finally, under severe compound degradation (JPEG $Q=50$ and Gaussian blur; Fig.~\ref{fig:robustness}f), FMGAN consistently achieves peak BA, verifying its robust extraction performance under realistic conditions.

The significantly higher BA values achieved by HiDDeN and FMGAN compared to their NN counterparts confirm that incorporating a noise subnetwork is critical for enhancing robustness against adversarial distortions. Moreover, the superior performance of FMGAN-NN over HiDDeN-NN further demonstrates that the proposed feature fusion mechanism contributes to stable watermark extraction even in the absence of explicit noise-aware training.

\subsubsection{Utility Impact on the Original FL Task}
\label{subsubsec:utility-impact}

\begin{table}[t]
	\centering
	\caption{Utility impact of the proposed scheme on the original FL task.
	The \textbf{Watermark} column reports test accuracy when clients train on
	watermarked images; the remaining columns report accuracy after enabling
	each private validation component.}
	\label{tab:utility-impact}
	\setlength{\tabcolsep}{4pt}
	\renewcommand{\arraystretch}{1.12}
	\small
	\begin{tabular}{@{}lcccccc@{}}
		\toprule
		\multirow{2}{*}{\textbf{Dataset}}
		& \multirow{2}{*}{\textbf{Orig. FL}}
		& \multicolumn{5}{c}{\textbf{Proposed scheme}} \\
		\cmidrule(l){3-7}
		& 
		& \textbf{WM}
		& \textbf{\(\ZKTwo\)}
		& \textbf{\(\ZKInf\)}
		& \textbf{\(\ZKCos\)}
		& \textbf{Full} \\
		\midrule
		MNIST-0.5
		& 96.3\%
		& 96.1\%
		& 96.2\%
		& 96.2\%
		& 96.1\%
		& 96.1\% \\
		CIFAR-10
		& 83.8\%
		& 83.6\%
		& 83.7\%
		& 83.6\%
		& 83.5\%
		& 83.4\% \\
		\bottomrule
	\end{tabular}
\end{table}

We verify that the proposed scheme does not degrade the utility of the original FL task. 
By comparing the global model accuracy when clients are trained on clean and watermarked images (Table~\ref{tab:utility-impact}), we observe that the watermarked setting achieves an accuracy of 83.6\%, only 0.2 percentage points lower than the clean baseline of 83.8\%.
These results confirm that the high-fidelity FMGAN embedding preserves downstream training performance.

Valid client updates are still aggregated under the same training objective, while the additional constraints only filter updates that violate the public validation parameters. Hence, under benign training, the gap between \emph{Original FL} (the baseline without private validation) and the
\emph{Full scheme} in Table~\ref{tab:utility-impact} stems mainly from fixed-point representation and the exclusion of updates that fail these parameters.
The full scheme changes the accuracy by only \(0.2\) percentage points on MNIST-0.5 and \(0.4\) percentage points on CIFAR-10, confirming its small impact on the original learning task.

The scheme also preserves the robustness benefit of the validation rules under attack: on CIFAR-10 under a Trim attack, accuracy collapses from \(83.8\%\) to \(19.5\%\) without validation, but recovers to \(80.1\%\) once \(\ZKTwo\), \(\ZKInf\), and \(\ZKCos\) are enabled.
Thus the scheme incurs only a small benign-case cost while retaining norm- and direction-based robustness.
In practice, \(\ZKTwo\) and \(\ZKInf\) suffice when only magnitude control is needed to limit update amplification and model replacement, while \(\ZKCos\) adds directional consistency against the public cosine-similarity threshold.

\subsection{Experimental Discussions and Limitations}
\label{sec:exp_summary}

Experimental results validate our dual-layer framework at both component and system levels.
At the data layer, the keyed PAM watermarking mechanism ensures strong imperceptibility and robustness with negligible overhead, leveraging HMAC-SHA256 to align authenticity with Kerckhoffs' principle.
At the computation layer, LZKSA secure aggregation enforces gradient-level integrity; its zero-knowledge proofs (validating keys, norm bounds, and cosine similarity) successfully filter out malicious scaling and directional updates while preserving privacy and scaling linearly with client size.
Finally, system-level evaluations confirm that this joint pipeline robustly mitigates composite cross-layer threats—including concurrent dataset replacement, forged data injection, noise perturbations, and protocol abuse—without sacrificing aggregation accuracy or watermark extractability.

Consistent with the generation-unforgeability scope claimed above, two properties remain for future work:
an indirect physical anchor coupling dependency and vulnerability to watermark-forgery (replay) attacks.
First, the current watermarking mechanism relies on PAM as an indirect reference (a "soft link") to the ground-truth anchor fabric rather than embedding the real-time ENF signal directly into the host images (a "hard link"), which introduces an abstract dependency layer between the media and its environmental anchor.
Second, while PAM provides broad context-binding for image datasets, it lacks intrinsic content-binding capabilities and cannot independently prevent watermark forgery; an adversary could potentially extract valid watermarks from leaked assets and maliciously transplant them onto unauthorized images, creating severe attribution vulnerabilities.
To mitigate these constraints, our future research will focus on developing a unified provenance token structure that fuses context-binding metadata (such as direct ENF signals or alternative environmental anchors) with content-binding data (such as image perceptual hashes or semantic features) to support completely self-contained, cryptographic verification.



Additionally, the current watermark robustness evaluation primarily focuses on common image distortions and noise-based perturbations. Although these represent realistic degradation scenarios, more adaptive and optimization-based removal attacks could be explored to further stress-test watermark resilience. 
Moreover, our evaluation focuses on image-based federated learning scenarios. Extending the keyed watermarking mechanism to other data modalities, such as text, audio, or multimodal settings, remains an important direction for future research.

\section{Conclusion and Future Work}
\label{sec:Conclusion}

This paper presented a dual-layer security framework for trustworthy federated learning. At the data layer, we designed a Kerckhoffs-compliant keyed PAM watermarking scheme leveraging FMGAN with Mamba-guided attention to achieve robust, imperceptible ownership validation and leakage tracing. At the computation layer, we integrated a lattice-based zero-knowledge secure aggregation protocol over RLWE commitments, enforcing gradient-level key correctness, norm bounds, and directional constraints under post-quantum security. Comprehensive system evaluations confirm that this joint pipeline robustly mitigates cross-layer threats, including dataset substitution, forged data injection, gradient scaling, and protocol abuse.
Future work will focus on studying adaptive watermark removal attacks, such as optimization- or learning-based strategies, to further strengthen resilience under fully adaptive threat models.
Additionally, we plan to extend the keyed physical anchor watermarking scheme to other modalities (e.g., text, audio, or multimodal data) is another important direction.
Furthermore, optimizing zero-knowledge verification for extremely large-scale federated deployments may further improve latency efficiency.

\bibliographystyle{IEEEtranS}
\bibliography{references}

\end{document}